\documentclass[referee]{aa} 

\usepackage{graphicx}

\def\a209{$\mathrm{A\,209}$} 
\def\ac118{$\mathrm{AC\,118}$}
\def\cl0048{$\mathrm{EIS\,0048}$} 
\def\re{$\mathrm{r_e}$}

\def\ns{$\mathrm{n}$}
\def\lesssim{\widetilde{<}}
\begin{document}

\titlerunning{Evolution of UV -- NIR Structural Properties }

   \title{Evolution of UV -- NIR Structural Properties of Cluster
Galaxies\thanks{Based on observations collected at
European Southern Observatory. Figures C.1, C.2, and C.3 are only available 
in electronic form at http://www.edpsciences.org.}}

    \subtitle{ }

\author{F. La Barbera \inst{1} 
\and
G. Busarello  \inst{1} 
\and 
M. Massarotti \inst{1} 
\and 
P. Merluzzi   \inst{1}
\and
A. Mercurio   \inst{2} 
   }

\offprints{F. La Barbera}

\institute{ 
 I.N.A.F., Istituto Nazionale di Astrofisica 
 Osservatorio Astronomico di Capodimonte, 
 Via Moiariello 16, I-80131 Napoli \\ 
 email: labarber@na.astro.it 
 \and 
 Universit\`a degli Studi di Trieste, Department of Astronomy, Trieste \\
}

\date{Received ; accepted }

\abstract{ We study the structure and the internal colour gradients of
  cluster galaxies from UV to NIR restframe, in the redshift range
  $\mathrm{z=0.21-0.64}$. Structural parameters (half light radius
  \re, mean surface brightness $\mathrm{< \! \mu \! >_e}$ and Sersic
  index \ns) are derived for galaxies in the clusters \a209 \, at
  $\mathrm{z = 0.21}$ and \cl0048 at $\mathrm{z = 0.64}$. This data
  set, together with previous data for the cluster \ac118 at
  $\mathrm{z=0.31}$, constitutes the first large ($\mathrm{N \sim
  270}$) sample of cluster galaxies whose internal structure in UV,
  optical (OPT) and NIR (U-, V- and H-band restframe) can be
  investigated up to a look-back time of $\mathrm{\sim 6~Gyr}$
  ($\Omega_m=0.3$, $\Omega_\Lambda=0.7$ and $\mathrm{ H_0=
  70~Km~s^{-1} Mpc^{-1} }$). Galaxies are classified as spheroids or
  disks according to the shape of the light profile, and the evolution
  of the two populations are investigated separately.  On average,
  both spheroids and disks are more concentrated at longer
  wavelengths: the galaxy sizes become smaller from UV to NIR, while
  Sersic indices increase. This trend shows an evolution in disks,
  where the mean ratio of optical to NIR Sersic indices decreases from
  $\mathrm{z=0.31}$ to $\mathrm{z=0.64}$.  Colour gradients are on
  average negative at all redshifts and are stronger in disks than in
  spheroids. But while for spheroids both $\mathrm{grad(UV-OPT)}$ and
  $\mathrm{grad(OPT-NIR)}$ are only weakly dependent on redshift, the
  optical-NIR gradients of disks become significantly smaller at
  $\mathrm{z=0.64}$. Colour gradients and central colours are compared
  with models of metallicity, age, and dust extinction
  gradients. Metallicity turns out to be the primary driver of colour
  gradients in spheroids, the age gradient being constrained to be
  smaller than $\sim 25\%$. For disks, two kinds of models fit the
  present data: (i) age gradients (in the range
  $\left[30,50\right]\%$) with significant dust extinction, and (ii)
  `pure' dust models, in which the gradients of colour excess are a
  factor of two higher in \cl0048 than in the other clusters.  Since
  colour gradients of disks seem not to correlate significantly with
  inclination, we argue that age gradient models could represent a
  more likely explanation of the present data, in agreement with what
  expected on the basis of hierarchical merging scenarios.

   }

   \maketitle
%


\section{Introduction}
Studies of spectral indices and broadband colours have demonstrated
that stellar populations in galaxies are not homogeneous, but are
characterized by a spatial variation in their physical properties.

Most of nearby early-type (E) galaxies have negative colour gradients,
becoming bluer toward the periphery (e.g. Boroson et al.~\cite{BTS83};
Franx et al.~\cite{FIH89}; Peletier et al.~\cite{PDI90},~\cite{PVJ90},
hereafter PDI90 and PVJ90, respectively; Michard~\cite{Mic99}). The
presence of radial changes in colour may be interpreted in terms of
both age and metallicity, due to the well known degeneracy between the
two properties (Worthey et al.~\cite{WTF96}). The most effective way
to investigate the origins of colour gradients is to study their
evolution with look-back time, since age and metallicity models give
very different predictions at increasing redshifts (Kodama \&
Arimoto~\cite{KoA97}).  By adopting this approach, Saglia et
al.~(\cite{SMG00}) and Tamura \& Ohta~(\cite{TaO00}) analyzed the
colour profile of Es in distant clusters by using HST optical
photometry. They found that galaxies show moderate colour gradients
also at intermediate redshifts, and claimed, therefore, that
metallicity is the primary factor driving the observed gradients. The
same conclusion was drawn by Tamura et al.~(\cite{TKA00}) and by
Hinkley \& Im~(\cite{HII01}) for early-type galaxies in the field.  On
the other hand, there are also some indications that the population of
spheroids at intermediate redshifts is not homogeneous with respect to
the radial properties of the stellar populations. Recently,
Bartholomew et al.~(\cite{BRG01}) showed that cluster K+A galaxies at
intermediate redshifts could have somewhat bluer colour gradients with
respect to normal early-types, while Menanteau et al.~(\cite{MAE01})
found positive colour gradients for spheroids in the Hubble Deep
Fields, concluding that a significant fraction of galaxies have
experienced recent star formation localized in the center.

Various attempts have been made to interpret the existence of colour
gradients of early-type galaxies in the framework of different
evolutionary models.  The presence of metallicity gradients is
naturally expected within the monolithic collapse scenarios
(Larson~\cite{Lar74}), as a consequence of the later onset of the
galactic wind in the inner region of the galaxy, but it can also be
accommodated within the framework of galaxy formation via hierarchical
merging (White~\cite{Whi80}; Mihos \& Hernquist~\cite{MiH94}).
Positive colour gradients are not predicted by the monolithic collapse
while they can be explained as a consequence of the centrally peaked
star formation produced in merging remnants (Mihos \&
Hernquist~\cite{MiH94},~\cite{MiH96}).  Recently, it has been shown
that colour gradients of spheroids can also be explained within the
chemodynamical model for evolution of elliptical galaxies (see Fria\c
ca \& Terlevich~\cite{FrT01} and references therein).

The origin of colour gradients is crucial to understand the mechanisms
underlying the formation and evolution of late-type galaxies. Colour
gradients in nearby disks have been extensively investigated by de
Jong~(\cite{deJ96}) using optical and near-infrared broadband
photometry. He found that disk galaxies have more metal rich and older
stellar populations in the center, and that, therefore, both age and
metallicity drive the observed radial colour profiles. The presence of
age gradients is also supported by the increase of the disk
scale-lengths measured by the $\mathrm{H_{\alpha}}$ flux with respect
to the optical (Ryder \& Dopita~\cite{RyD94}).  Further investigations
have been performed by Gadotti \& dos Anjos~(\cite{GAD01}), who
studied a large sample of late-tape spirals in the field by UV --
optical photometry.  They found that most of the galaxies ($85 \%$)
have negative or null colour gradients, and suggested that colour
gradients are more sensitive to age than to metallicity of stellar
populations. Disks in the cluster environment have been studied with
NIR (J -- K) data by Moriondo et al.~(\cite{MBC01}), who found very
weak or null colour gradients.

The presence of age gradients is a robust prediction of hierarchical
models, in which spiral bulges contain older stars than their
associated disks, which form by subsequent accretion.  Evidence in
favour of the presence of age gradients for field spirals at
intermediate redshifts has been found by Abraham et
al.~(\cite{AEF99}).

A crucial role in the interpretation of colour gradients can be played
by dust absorption. Although the effects of extinction remain still
unclear and substantially unresolved (e.g. Tamura et al.~\cite{TKA00};
de Jong~\cite{deJ96}; Gadotti \& dos Anjos~\cite{GAD01}), different
studies have shown that dust absorption could have an important effect
in the internal colour distributions of galaxies (e.g. Wise \&
Silva~\cite{WiS96}; Silva \& Elston~\cite{SiE94}). In particular, for
what concerns disk dominated galaxies, it has been proved that
internal extinction strongly affects the colour profile of late-type
spirals (e.g. Peletier et al.~\cite{PVM95}).  For these reasons, in
the present work we consider simple extinction gradient models for
disk galaxies.

An effective tool to analyze the internal light distribution in
galaxies is the comparison of structural parameters (like the half
light radius and the Sersic index) at different wavelengths.  This
approach allows large samples of galaxies to be studied with a broad
photometric coverage.  By using ground-based data, La Barbera et
al.~(\cite{LBM02}, hereafter LBM02) obtained the first large sample
($\mathrm{N =94}$) of both optical and NIR structural parameters for
cluster galaxies at $\mathrm{z \sim 0.3}$. They suggested that both
age and metallicity could drive the observed colour gradients in the
cluster populations. In the present work, we re-analyze this subject
by using a larger sample ($\mathrm{N \sim 270}$) of cluster galaxies
at different redshifts ($z = 0.21, 0.31, 0.64$) with a broader
photometric baseline: from UV to NIR restframe. We estimate colour
gradients for both disks and spheroids, with the aim of following the
properties of both cluster populations down to $\mathrm{z = 0.64}$.

The layout of the paper is the following. In Sect.~\ref{SECDATA} we
describe the samples used for the present analysis. The derivation of
structural parameters is outlined in Sect.~\ref{SECSF}, while
Sect.~\ref{SECUKPAR} deals with the comparison of UV -- NIR parameters
at the different redshifts.  The distributions of colour gradients are
analyzed in Sect.~\ref{SECCGRAD}, and their evolution with redshift is
compared with predictions of metallicity, age, and extinction models
in Sect.~\ref{SECECGRAD}. Summary and conclusions are drawn in
Sect.~\ref{SECCONC}. In the following, we will assume the cosmology
$\Omega_m=0.3$, $\Omega_\Lambda=0.7$ and $\mathrm{ H_0= 70~Km~s^{-1}
Mpc^{-1} }$. With these parameters, the age of the universe is
$\mathrm{\sim13.5~Gyr}$, and the redshifts $\mathrm{z=0.21, z=0.31}$
and $\mathrm{z=0.64}$ correspond to look-back times of $\mathrm{\sim
2.5, 3.5}$ and $\mathrm{6~Gyr}$, respectively.

\section{The data}
\label{SECDATA}
The data relevant for the present study have been collected at the ESO
New Technology Telescope and at the ESO Very Large Telescope. They
include images in B and R bands for the cluster \a209 at
$\mathrm{z=0.21}$ ($\mathrm{RA_{2000}=01 \!: \! 31 \!: \! 57.5}$,
$\mathrm{DEC_{2000}=-13 \!: \! 34 \! : \! 35}$), in R and K bands for
the cluster \ac118 at $\mathrm{z=0.31}$ ($\mathrm{RA_{2000}=00 \!: \!
14 \!: \! 19.5}$, $\mathrm{DEC_{2000}=-30 \!: \! 23 \! : \! 19}$), and
in V, I (high resolution, HR) and K bands for the cluster \cl0048 at
$\mathrm{z=0.64}$ ($\mathrm{RA_{2000}=00 \!: \!  48 \!: \! 31.6}$,
$\mathrm{DEC_{2000}=-29 \!: \! 43 \! : \! 18.0}$). \a209 and \ac118
are rich, massive clusters of galaxies, with velocity dispersions of
$\mathrm{1394 \pm 94~km~s^{-1}}$ (Mercurio et al.~\cite{MGB03},
hereafter MGB03) and $\mathrm{1950 \pm 334~km~s^{-1}}$ (Wu et
al.~\cite{WXF99}), respectively, and with high X-ray luminosity
($\mathrm{L_X \sim 13.8~10^{44}~erg~s^{-1}}$, see Ebeling et
al.~\cite{EVB96}, and $\mathrm{L_X=62.44 \pm
14.41~10^{44}~ergs~s^{-1}}$, see Wu et al.~\cite{WXF99},
respectively). Their velocity field and spatial structure suggest the
presence of substructures, with a complex dynamical scenario (see
MGB03 and Andreon~\cite{A01}, respectively). \cl0048 is a cluster of
galaxies having a velocity dispersion of $\mathrm{ \sim 720~km~s^{-1}
}$ (Lobo et al.~\cite{LSD02}), which is typical for clusters at the
same redshift, and a spatial structure which shows evident signs of
substructures (see La Barbera et al.~\cite{LMI03}). A ROSAT PSPC
image, which serendipitously contains the cluster field, indicates the
presence of X-ray emission.

The reduction of the photometric data of A\,209 and AC\,118, and that
of the V- and K-band images for EIS\,0048 is described in Mercurio et
al.~(2003b, MMM03), Busarello et al.~(2002) and Andreon~(2001), and La
Barbera et al.~(2003, hereafter LMI03), respectively. The reduction of
the I(HR)-band image for \cl0048 \, is described in Appendix A.
Structural parameters have been derived for galaxies of \ac118 in
LBM02\footnote{ With respect to LBM02, we excluded two galaxies whose
light profiles turned out to be strongly affected by nearby objects.},
while the surface photometry for \a209 \, and \cl0048 \, is presented
in the next section.  The relevant information on the samples are
summarized in Table~\ref{DATA}.

\begin{table*}
\caption[]{Relevant information on the samples. Column 1: cluster
identification. Column 2: redshift. Column 3: wavebands.  Column 4:
instruments used for the observation.  Column 5: pixel scale relative
to each waveband in arcsec/pixel.  Column 6: FWHM (arcsec) of the
seeing for each waveband.  Column 7: number of spheroids.  Column 8:
number of disks. For \cl0048 two values are reported, since the sample
with K-band data covers a smaller area.  Column 9: completeness
magnitudes.  Column 10: selection used to assign cluster membership,
$\mathrm{z_p}$ and $\mathrm{z_s}$ denote photometric and spectroscopic
redshifts, respectively.  Column 11: references for details on data
reduction and on the definition of cluster membership. The following
abbreviations are used: Mercurio et al.~(\cite{MGB03}, MGB03),
Mercurio et al.~(\cite{MMM03}, MMM03), Busarello et al.~(\cite{BML02},
BML02), Andreon~(2001, A01), La Barbera et al.~(\cite{LMI03},
LMI03). Note that one out of the $\mathrm{N=20}$ disks with OPT and
NIR data in \cl0048 does not have UV structural parameters, and
therefore the total number of disks at $\mathrm{z=0.64}$ is $38$.}
\label{DATA}
\scriptsize
\begin{center}
$$
\hspace{-0.8cm}
\begin{array}{c|c|c|c|c|c|c|c|c|c|c}
\hline
\noalign{\smallskip}
 \mathrm{CLUSTER} &  {\rm z}  &  \mathrm{Wavebands} & \mathrm{Instr.} & \mathrm{Scale} & \mathrm{Seeing} & \mathrm{N_{SPH}} & \mathrm{N_D} & \mathrm{M_c} & \mathrm{Sel.} & \mathrm{Ref.} \\
  (1) &  (2)  &  (3) & (4) & (5) & (6) & (7) & (8) & (9) & (10) & (11) \\
\noalign{\smallskip}
\hline
\noalign{\smallskip}
 \mathrm{A\,209}   & 0.21 & \mathrm{B, R}    & \mathrm{EMMI} & 0.27, 0.27     & 0.8, 0.9 & 64      & 26 & \mathrm{ R=20} & \mathrm{z_s} & \mathrm{MGB03, MMM03} \\
 \mathrm{AC\,118}  & 0.31 & \mathrm{R, K}    & \mathrm{EMMI, SOFI} & 0.27, 0.29     & 1.0, 0.8 & 76      & 15 & \mathrm{ R=21} & \mathrm{z_p} & \mathrm{BML02, A01} \\
 \mathrm{EIS\,0048} & 0.64 & \mathrm{V, I, K} & \mathrm{FORS2, ISAAC}  & 0.2, 0.1, 0.15 & 0.6, 0.35, 0.45    &  54,32   & 37,20 & \mathrm{I=22} & \mathrm{z_p} & \mathrm{LMI03} \\
\noalign{\smallskip}
\hline
\end{array}
$$
\end{center}
\end{table*}

\begin{figure}
\begin{center}
\resizebox{12cm}{6cm}{\includegraphics{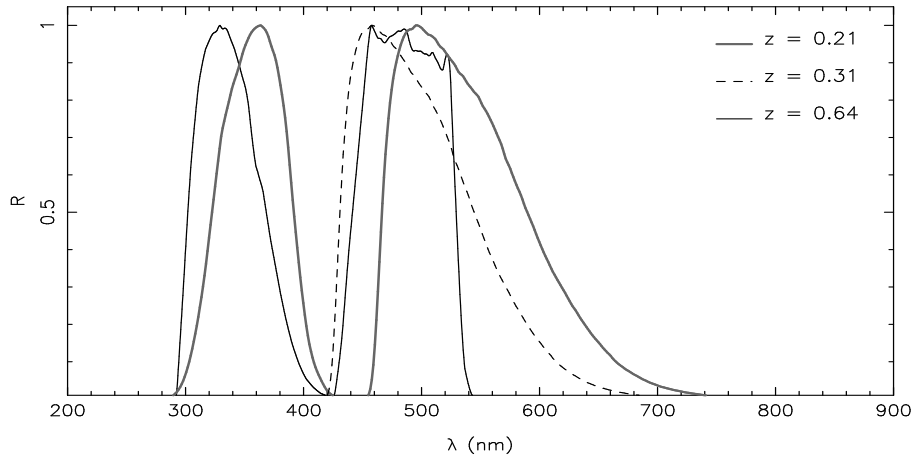}}
\end{center}
\caption[]{ Restframe filter response of the optical photometric
wavebands for \a209, \ac118, and \cl0048. The curves are normalized to
1 at the maximum and are drawn by different types of lines as shown in
the upper -- right.
\label{FILT}
}
\end{figure}

In Fig.~\ref{FILT}, we show the response curves of the optical filters
in the restframe for each cluster. We see that the R band for \a209
and \ac118, and the I band for \cl0048 cover approximately the same
restframe spectral interval, with a central wavelength ranging between
$ \mathrm{ \sim 4800 \stackrel{ \circ }{A} }$ and $ \mathrm{ \sim 5200
\stackrel{ \circ }{A} }$. These wavebands are close to the V-band
restframe and will be indicated in the following as optical (OPT)
wavebands. The B filter at $\mathrm{z = 0.21}$ and the V filter at
$\mathrm{z = 0.64}$ have central wavelengths in the range $3200$ --
$3600 \mathrm{ \stackrel{ \circ }{A} }$, and correspond therefore to U
restframe. In the following, they will be indicated as UV wavebands.
For what concerns the K band (hereafter NIR), it covers approximately
H and J restframe at $\mathrm{z=0.31}$ and at $\mathrm{ z=0.64 }$,
respectively. We note that colour gradients and variations in the
structural parameters between J and K bands are expected to be very
small (see Silva \& Elston~\cite{SiE94}) and therefore the differences
of the restframe corresponding to the K bands at $z=0.31$ and $z=0.64$
are negligible for the present analysis.  The data of \a209 and \ac118
allow UV -- OPT and OPT -- NIR properties of cluster galaxies at $ 0.2
< \mathrm{z} < 0.3$ to be studied, while the same wavelength range is
covered at $\mathrm{z=0.64}$ by V, I and K bands. Since K corrections
do not affect differences in colour, the UV-OPT and OPT-NIR colour
gradients correspond closely to U-V and V-H gradients restframe.  We
point out that the present samples constitute the first large data set
of cluster galaxies with structural parameters measured from UV to NIR
at intermediate redshifts.

For each cluster we study separately the properties of disks and
spheroids. The separation between the two classes is performed on the
basis of the Sersic index $\mathrm{n}$ in the optical, since the
optical parameters are common to each cluster sample. We define as
spheroids the objects with $\mathrm{n > 2}$, and as disks the
remaining galaxies. Following van Dokkum et al.~(\cite{vDF98}), this
criterion corresponds to distinguishing objects with a low bulge
fraction ($<20 \%$) from galaxies with a more prominent bulge
component. We note that the mean value\footnote{The mean value is
computed for \ac118 \, and \cl0048 \, galaxies, for which we obtained
$\mathrm{\overline{n}=1.2\pm0.1}$ and
$\mathrm{\overline{n}=1.0\pm0.05}$, respectively. For \a209 \, we
found a higher value, $\mathrm{\overline{n}=1.5\pm0.1}$, due to the
presence of a larger fraction of galaxies with $\mathrm{n \rightarrow
2}$. We found, however, that considering as disks only the galaxies
with $\mathrm{n<1.3}$ in \a209 \, does not change the estimates
(e.g. mean colour gradients) given throughout the paper. } of
$\mathrm{n}$ for disks is $1.10\pm0.06$. By using numerical
simulations, we estimated that this value describes galaxies with a
very low bulge fraction ($\mathrm{L_B/L_{TOT}} \sim 0.07$). Therefore,
although central bulges can affect the properties of some disks, their
effect on the bulk properties of disks can be neglected.  We also note
that, taking into account luminosity distances and K+E corrections,
the completeness magnitudes in Table~\ref{DATA} correspond
approximately to the same absolute magnitude limits\footnote{Colour
gradients of nearby galaxies do not correlate with luminosity (see
PDI90), and, therefore, slightly different magnitude limits should not
affect studies of colour gradients.}: $ \mathrm{M_V = -19.70\pm0.06}$
at $\mathrm{z = 0.21}$, $ \mathrm{M_V = -19.7\pm0.1}$ at $\mathrm{z =
0.31}$, and $ \mathrm{M_V \sim -20.0\pm0.2}$ at $\mathrm{z=0.64}$. K
and evolution corrections were obtained by the GISSEL00 synthesis code
(Bruzual \& Charlot~\cite{BrC93}), with spectral models having an
exponential star formation rate, and the uncertainties on
$\mathrm{M_V}$ were estimated by computing E corrections for different
models, exploring a wide range of stellar population parameters
(i.e. metallicity, formation epoch, time scale of star formation).

\begin{table*}
\caption[]{Comparison of structural parameters of disks. 
}
\label{Dpar}
\begin{center}
$$
\begin{array}{c|cc|cc|cc|cc}
\hline
\noalign{\smallskip}
  & \multicolumn{4}{c|}{\mathrm{UV \! - \! OPT}} & \multicolumn{4}{c}{\mathrm{OPT \! - \! NIR}} \\
\mathrm{z} & \multicolumn{2}{c}{\mathrm{\Delta (\log r_e)}} & \multicolumn{2}{c|}{\mathrm{\Delta(\log n)}} & \multicolumn{2}{c}{\Delta (\log r_e) } & \multicolumn{2}{c}{\mathrm{\Delta(\log n)} } \\
  & {\rm mean} & {\rm std} & {\rm mean} & {\rm std} & {\rm mean} & {\rm std} & {\rm mean} & {\rm std} \\
\noalign{\smallskip}
\hline
\noalign{\smallskip}
{\rm 0.21} & 0.01  \pm 0.04  & 0.08 \pm 0.02 & -0.09 \pm 0.02 & 0.10 \pm 0.04 & ..... & ..... & ..... & ..... \\
{\rm 0.31} &  ..... &  ..... & ..... & ..... & 0.09 \pm 0.02 & 0.10 \pm 0.02 & -0.17 \pm 0.04 & 0.20 \pm 0.05 \\
{\rm 0.64} & 0.045 \pm 0.016 & 0.09 \pm 0.02 & -0.06 \pm 0.03 & 0.15 \pm 0.02 &  0.10 \pm 0.02 & 0.13 \pm 0.07 & -0.34 \pm 0.05 & 0.25 \pm 0.06 \\
\noalign{\smallskip}
\hline
\end{array}
$$
\end{center}
\end{table*}

\begin{table*}
\caption[]{Comparison of structural parameters of spheroids. 
}
\label{Spar}
\begin{center}
$$
\begin{array}{c|cc|cc|cc|cc}
\hline
\noalign{\smallskip}
  & \multicolumn{4}{c|}{\mathrm{UV \! - \! OPT}} & \multicolumn{4}{c}{\mathrm{OPT \! - \! NIR}} \\
\mathrm{z} & \multicolumn{2}{c}{\mathrm{\Delta (\log r_e)}} & \multicolumn{2}{c|}{\mathrm{\Delta( \log n)}} & \multicolumn{2}{c}{\Delta (\log r_e) } & \multicolumn{2}{c}{\mathrm{\Delta( \log n)} } \\
  & {\rm mean} & {\rm std} & {\rm mean} & {\rm std} & {\rm mean} & {\rm std} & {\rm mean} & {\rm std} \\
\noalign{\smallskip}
\hline
\noalign{\smallskip}
{\rm 0.21} & 0.050 \pm 0.025  & 0.12 \pm 0.02 & -0.02 \pm 0.07 & 0.19 \pm 0.03 & .... & ..... & ..... & ..... \\
{\rm 0.31} & .....  & ..... & ..... & ..... & 0.21 \pm 0.05 & 0.26 \pm 0.04 & -0.03 \pm 0.02 & 0.14 \pm 0.04 \\
{\rm 0.64} & 0.12 \pm 0.03 & 0.24 \pm 0.03 & 0.05 \pm 0.06 & 0.28 \pm 0.04 & 0.080 \pm 0.015 & 0.14 \pm 0.04 & -0.19 \pm 0.05 & 0.20 \pm 0.20 \\
\noalign{\smallskip}
\hline
\end{array}
$$
\end{center}
\end{table*}

\section{Derivation of structural parameters}
\label{SECSF}
Structural parameters were derived for \a209 \, and \cl0048 \, by
following the procedure described in LBM02.  We give here only a brief
outline of the method, while details can be found in that paper.

Galaxy images are fitted by the convolution of an accurate model of
the Point Spread Function (PSF) with a model of the galaxy light
distribution. In this work, we consider Sersic models:
\begin{equation}
\mathrm{
I(r) = I_0 \cdot exp(- b \cdot (r / r_e)^{1/n})
},
\end{equation}
where r is the equivalent radius, $\mathrm{r_e}$ is the effective
radius, $\mathrm{I_0}$ is the central surface brightness, $\mathrm{n}$
is the Sersic index, and b is a constant ($\mathrm{b \sim 2n-1/3}$,
see Caon et al.~\cite{CCD93}).  The fits were performed by
constructing an interactive mask for each galaxy, and by modelling
simultaneously overlapping objects.  The PSFs were modelled by a
multi-Gaussian expansion, taking into account possible variations with
the chip position. With respect to LBM02, we had also to consider that
the PSF of the FORS2 I-band images showed significant deviations from
the circular shape. Details can be found in Appendix B, where we
illustrate the PSF modelling for the I-band images of \cl0048.  The
fitted surface brightness profiles are shown in Appendix C.

For \ac118, the typical uncertainties on the structural parameters
were estimated in LBM02 by comparing the ground-based parameters with
measurements obtained by HST data.  The quoted uncertainties amount to
$\sim 38 \%$ in \re \, and $\sim 20 \%$ in \ns. For \a209, the errors
on \re \, and \, \ns \, were estimated by using numerical
simulations. We found $\mathrm{\delta(r_e)\sim20\%}$ and
$\mathrm{\delta(n)\sim13\%}$.  For \cl0048, due to the very good
quality of the images, the uncertainties on \re \, and \ns \, turned
out to be smaller.  By comparing measurements obtained for galaxies in
common between different pointings, we obtained $\delta r_e / r_e \sim
15 \%$ and $\delta n / n \sim 10 \%$.  The estimates of
$\mathrm{\delta(r_e)}$ and $\mathrm{\delta(n)}$ were used to derive
the uncertainties on the colour gradients (see Sect.~\ref{SECCGRAD}).

\section{Structure of galaxies from UV to NIR}
\label{SECUKPAR}

\begin{figure}
\begin{center}
\resizebox{9cm}{9cm}{\includegraphics{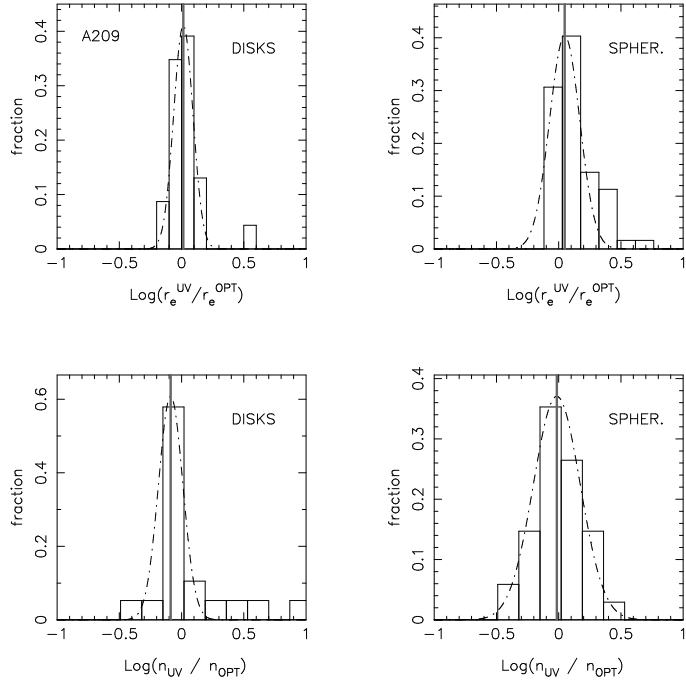}}
\end{center}
\caption[]{Comparison of UV and optical structural parameters of
spheroids (left panels) and disks (right panels) for the cluster
\a209. Upper and lower panels refer to the effective radii and Sersic
indices, respectively. The vertical grey line marks the mean of the
Gaussian fit (dot-dashed line) to the distribution.
\label{021par}
}
\end{figure}

\begin{figure}
\begin{center}
\resizebox{9cm}{9cm}{\includegraphics{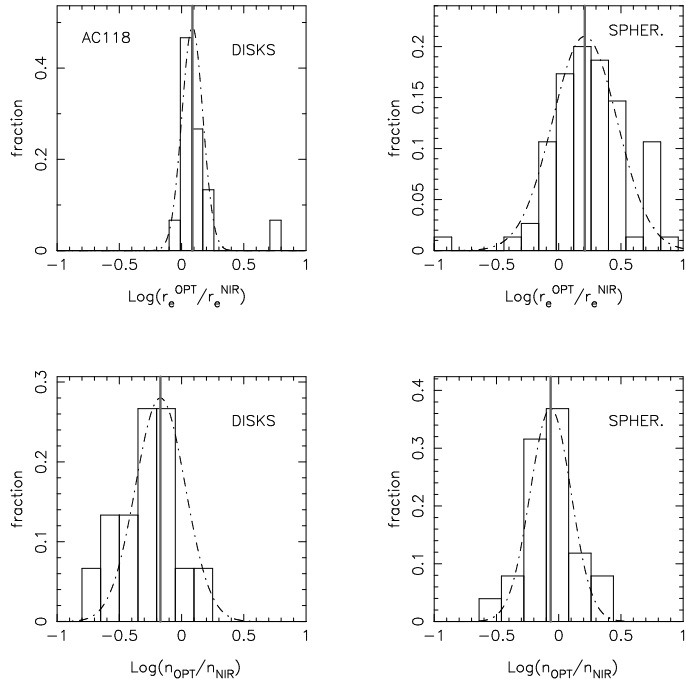}}
\end{center}
\caption[]{The same of Fig.~\ref{021par} for optical and NIR
structural parameters of galaxies in \ac118.
\label{031par}
}
\end{figure}

\begin{figure}
\begin{center}
\resizebox{9cm}{9cm}{\includegraphics{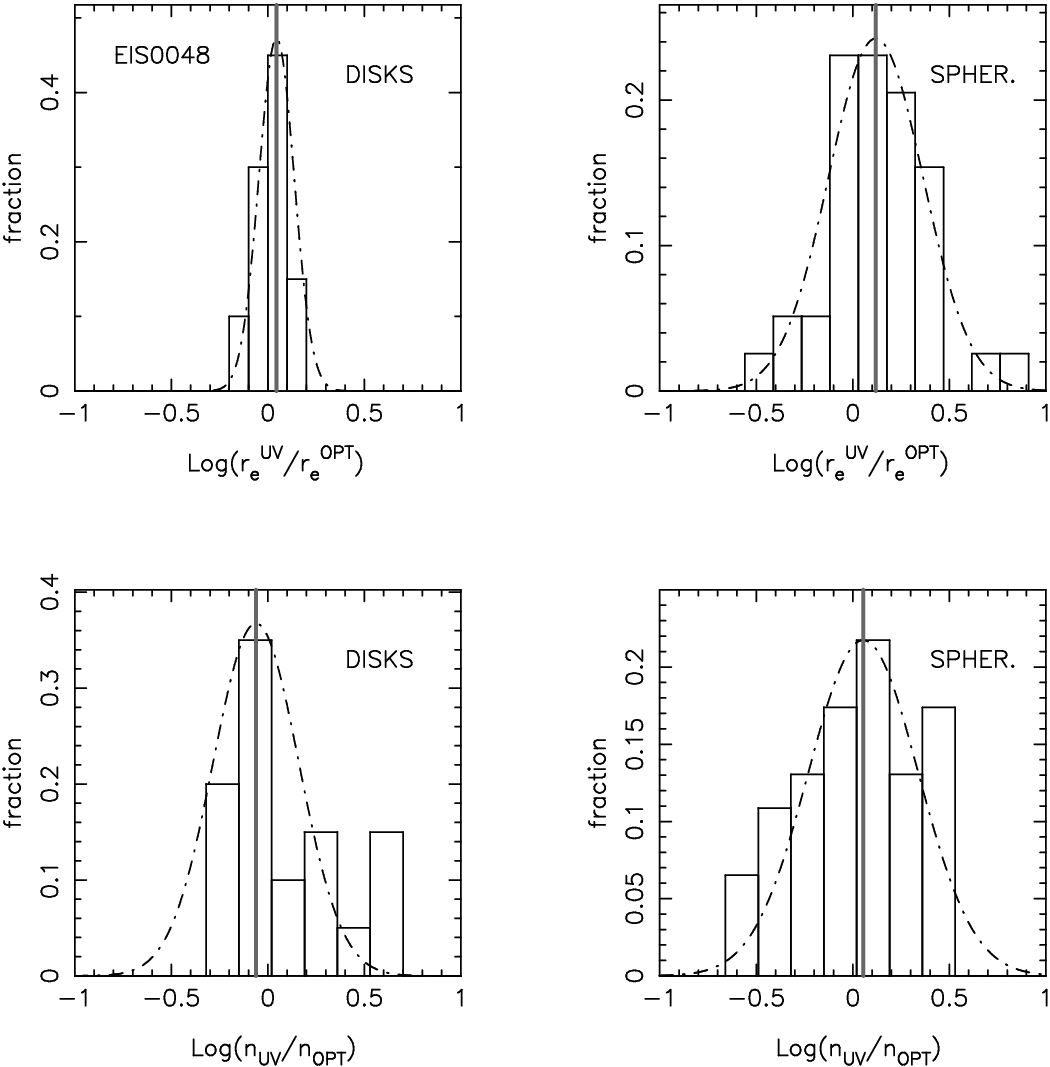}}
\end{center}
\caption[]{Same of Fig.~\ref{021par} for the cluster \cl0048. 
\label{064VIpar}
}
\end{figure}

\begin{figure}
\begin{center}
\resizebox{9cm}{9cm}{\includegraphics{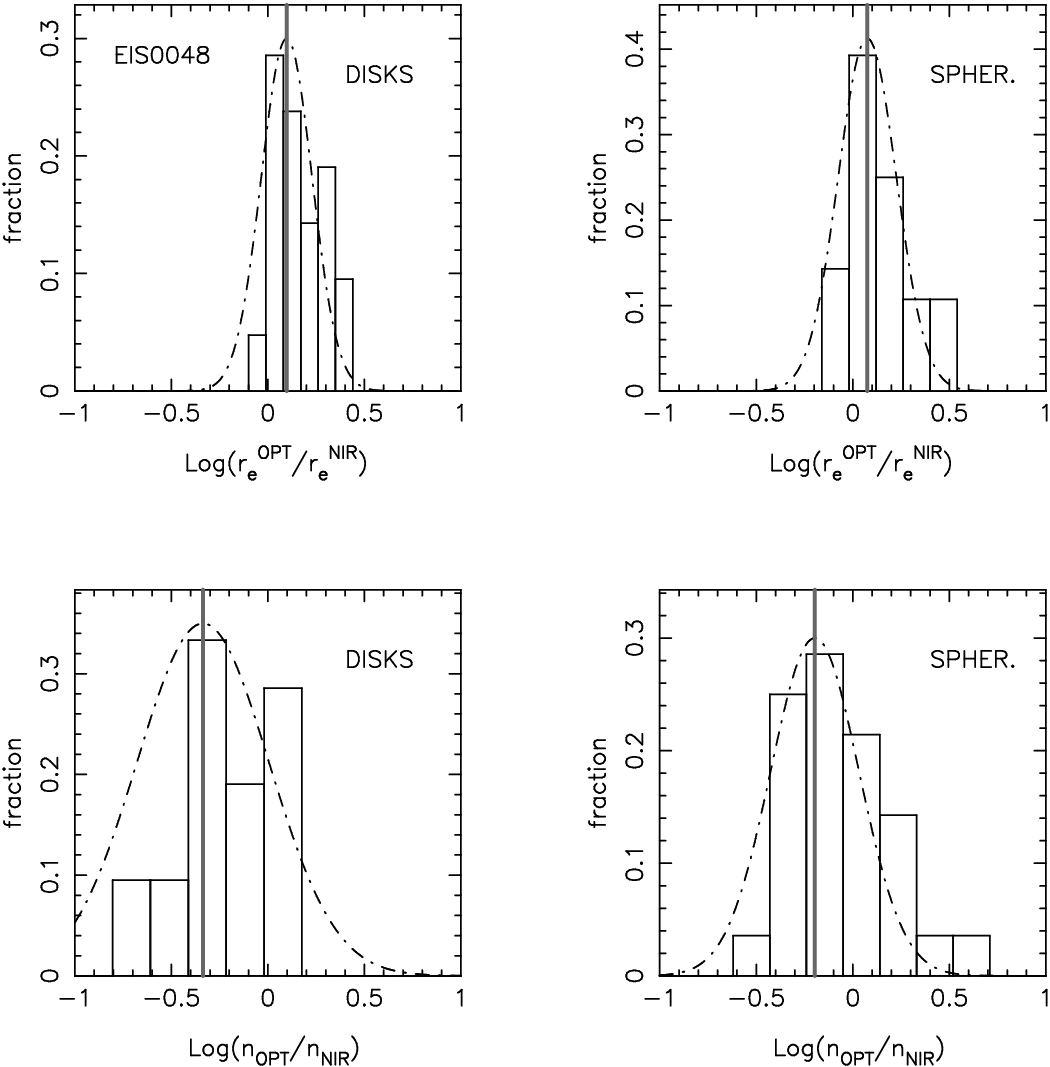}}
\end{center}
\caption[]{Same of Fig.~\ref{031par} for the cluster \cl0048.
\label{064IKpar}
}
\end{figure}

The shape of the light profile for the Sersic models can be fully
characterized in terms of the effective radius and the Sersic index.
The differences of \re \, and \ns \, between the various wavebands are
shown in Figs.~\ref{021par}, ~\ref{031par},~\ref{064VIpar}
and~\ref{064IKpar}. Differences are always computed between shorter
and longer wavelengths.  Figs.~\ref{021par} \, and~\ref{064VIpar} \,
show the UV -- OPT comparison for \a209 \, and \cl0048, respectively,
while Figs.~\ref{031par} and~\ref{064IKpar} show the OPT -- NIR
parameters for \ac118 \, and \cl0048, respectively. In each figure,
left and right panels refer to disks and spheroids, while upper and
lower panels refer to \re \, and \ns, respectively. In order to
analyze the overall properties of each distribution, we derived the
corresponding mean value and standard deviation by applying the
bi-weight statistics (e.g. Beers et al.~\cite{BFG90}), that has the
advantage to minimize the effect of outliers in the distributions.
The values and the corresponding uncertainties, obtained by the
bootstrap method, are shown in Tables~\ref{Dpar} and~\ref{Spar} for
disks and spheroids, respectively.

\subsection{Disks}
\label{PARDISK}
By looking at Table~\ref{Dpar}, we see that disk galaxies tend to have
more flat and extended profiles in the UV than in the optical. We
find, in fact, $\mathrm{n^{UV} < n^{OPT}}$ and $\mathrm{r_e^{UV} >
r_e^{OPT}}$ both for \a209 \, and for \cl0048, although the mean value
of $\mathrm{\Delta(\log r_e)}$ at $\mathrm{z=0.21}$ is not
significantly different from zero within the corresponding
uncertainty.  The mean value of $\mathrm{n^{UV} / n^{OPT}}$ is $\sim
0.8$ at $\mathrm{z = 0.21}$ and $\sim 0.87$ at $\mathrm{z = 0.64}$,
while the value of $\mathrm{r_e^{UV} / r_e^{OPT}}$ ranges from $\sim
1.02 $ for \a209 \, to $\sim 1.11$ for \cl0048.  We note that these
difference are within $1 \sigma $ and, therefore, are not very
significant. At $z = 0.64$, disks have larger dispersion in the
$\mathrm{ \log ( n^{UV} / n^{OPT} ) }$ diagram, with a standard
deviation that is $\sim 3/2$ of that at $z = 0.21$. The effect is
clearly seen by comparing Figs.~\ref{021par} and~\ref{064VIpar}, and
is particularly significant in view of the fact that the uncertainties
on the structural parameters of \cl0048 are smaller with respect to
those of \ac118 \, and \a209.

Disk galaxies in \ac118 \, and \cl0048 have, on average, ${\rm
r_e^{OPT} > r_e^{NIR}}$ and $\mathrm{ n^{OPT} < n^{NIR}}$ (see
Table~\ref{Dpar}), and show, therefore, the same trend observed in the
$\mathrm{UV}$ -- $\mathrm{OPT}$ comparison: at larger wavelengths, the
light profile becomes more concentrated. On average, the value of ${
\rm r_e^{OPT} / r_e^{NIR}}$ is $\sim 1.23$ at $\mathrm{z=0.21}$ and
$\sim 1.26$ at $\mathrm{z=0.64}$, and $\mathrm{ n^{OPT} / n^{NIR} }$
decreases from $\sim 0.7$ to $\sim 0.45 $.  While the behaviour of ${
\rm r_e^{OPT} / r_e^{NIR}}$ is similar to that found in the
$\mathrm{UV}$ -- $\mathrm{OPT}$, we see that the value of $\mathrm{
n^{OPT} / n^{NIR} }$ shows a much larger variation, with a strong
decrease for \cl0048.  We will further investigate this point in the
analysis of colour gradients (Sect.~\ref{CGDISK}). It is also
interesting to note that the distribution of $\mathrm{ \Delta \log
(n)}$ at $\mathrm{z=0.31}$ has a significant tail toward negative
values, showing that also for \ac118 \, there are some disks whose NIR
light is very peaked in the center with respect to the optical.

Finally, we point out that, since galaxies of \ac118 and \cl0048 are
selected by the photometric redshift technique, we cannot exclude that
the previous results can be affected by the contamination of the
cluster samples at $z = 0.64$ by field galaxies, whose properties,
also at low redshift, are known to be more disperse and heterogeneous
with respect to those of galaxies in the cluster environment.  The
role of field contamination will be further addressed in
Sect.~\ref{CGDISK}, in the comparison of colour gradients.

\subsection{Spheroids}
\label{PARSPHER}
By comparing $\mathrm{UV}$ and optical parameters we see that, on
average, spheroids have $\mathrm{r_e^{UV} > r_e^{OPT}}$. The mean
value of $\mathrm{r_e^{UV} / r_e^{OPT}}$ is $\sim 1.12 $ at
$\mathrm{z=0.21}$ and $\sim 1.32$ at $\mathrm{z=0.64}$, while
$\mathrm{n^{UV} / n^{OPT}}$ is consistent with zero at both redshifts
within the corresponding uncertainty. These results indicate that the
value of ${\rm r_e^{UV} / r_e^{OPT}}$ could increase slightly with
$\mathrm{z}$ while possible variations of $\mathrm{n^{UV} / n^{OPT}}$
are too small to be significant.  It is also interesting to note
that the dispersions of the two distributions are remarkably larger at
$\mathrm{z=0.64}$, ranging from $\sim 0.12$ to $\sim 0.24$ for
$\mathrm{\log r_e}$ \, and from $\sim 0.19$ to $\sim 0.28$ for
$\mathrm{\log n}$. This is similar to what found for the disks and
implies that the structural properties of cluster galaxies become more
heterogeneous at $\mathrm{z = 0.64}$. Since a low fraction of
spheroids at $z=0.64$ are expected to be field galaxies ($\sim 10$ --
$15 \%$, see Sect.~8.1 of LMI03), these results should not to be
affected by field contamination.

For what concerns the $\mathrm{OPT}$ -- $\mathrm{NIR}$ parameters, we
find that, on average, ${ \rm r_e^{OPT} > r_e^{NIR} }$ and $\mathrm{
n^{OPT} < n^{OPT}}$ for both \ac118 \, and \cl0048. The mean value of
$\mathrm{n^{OPT} / n^{NIR}}$ varies from $\sim 0.93$ at $\mathrm{z =
0.31}$ to $\sim 0.65$ at $\mathrm{z = 0.64}$, while ${ \rm r_e^{OPT} /
r_e^{NIR} }$ seems to be larger at $\mathrm{z = 0.31}$, ranging from
$\sim 1.20$ for \cl0048 \, to $\sim 1.6$ for \ac118.  This trend is
opposite to that found for disks and for spheroids in the
$\mathrm{UV}$ -- $\mathrm{OPT}$. As it is evident from the large
dispersion of the distribution in Fig.~\ref{031par} (upper -- right
panel), this effect could be not real but due to the uncertainties on
the estimated structural parameters at $\mathrm{z \sim 0.3}$. We note,
in fact, that also the standard deviation of the ${ \log( \rm
r_e^{OPT} / r_e^{NIR} ) }$ distribution shows a different trend with
$\mathrm{z}$ with respect to that found for the disks and for the
spheroids in the UV -- OPT: it seems to decrease from $\mathrm{z \sim
0.31}$ to $\mathrm{z \sim 0.64}$.  More NIR data at $\mathrm{ z < 0.6
}$ would be needed in order to further investigate the previous
trends.

\section{Colour gradients}
\label{SECCGRAD}
The dependence of structural parameters on the waveband carries
information on the properties of stellar populations at different
radii from the center of the galaxies. To investigate this subject, we
derived the internal $\mathrm{UV}$ -- $\mathrm{OPT}$ and
$\mathrm{OPT}$ -- $\mathrm{ NIR }$ colour gradients for \a209, \ac118
\, and \cl0048. We used the procedure described in LBM02: structural
parameters were used to derive the colour $\mu_2 - \mu_1$ of the
galaxy at an inner radius $\mathrm{r_i}$ and at an outer radius
$\mathrm{r_o}$. Here, the symbols $1$ and $2$ denote the two
wavebands. The colour gradient was obtained by the logarithmic
gradient of $\mu_2 - \mu_1$ (see Eq.~(4) in LBM02). In order to make a
direct comparison with previous works (PDI90, PVJ90), we adopted
$\mathrm{r_o = r_e}$ and $\mathrm{r_i = 0.1 \cdot r_e}$.

\begin{table}
\caption[]{${\rm UV - OPT}$ and ${\rm OPT - NIR}$ colour gradients and
central colours of disks.  }
\label{CGRAD_DISK}
$$
\begin{array}{ccccccc}
\hline
\noalign{\smallskip}
  \mathrm{z} & \multicolumn{2}{c|}{\mathrm{grad(UV \! - \! OPT)}} & \multicolumn{2}{c|}{\mathrm{grad(OPT \! - \! NIR)}} & \multicolumn{1}{c|}{\mathrm{UV-OPT}} & \mathrm{OPT-NIR} \\
   & {\rm mean} & {\rm std} & {\rm mean} & {\rm std} & {\rm mean} & {\rm mean} \\
\noalign{\smallskip}
\hline
\noalign{\smallskip}
 0.21 & -0.4 \pm 0.1 & 0.40 \pm 0.05 & ..... & ..... & 2.0\pm0.3& ..... \\
 0.31 & ..... & ..... & -0.68 \pm 0.12 & 0.87 \pm 0.30 & ..... & 3.80\pm0.13\\
 0.64 & -0.46 \pm 0.12 & 0.55 \pm 0.06 & -1.2 \pm 0.1 & 0.5 \pm 0.3 & 2.2\pm0.4 & 4.0 \pm 0.4\\
\noalign{\smallskip}
\hline
\end{array}
$$
\end{table}

\begin{table}
\caption[]{${\rm UV - OPT}$ and ${\rm OPT - NIR}$ colour gradients 
and central colours of spheroids.
}
\label{CGRAD_SPHER}
$$
\begin{array}{ccccccc}
\hline
\noalign{\smallskip}
  \mathrm{z} & \multicolumn{2}{c|}{\mathrm{grad(UV \! - \! OPT)}} & \multicolumn{2}{c|}{\mathrm{grad(OPT \! - \! NIR)}} & \multicolumn{1}{c|}{\mathrm{UV-OPT}} & \mathrm{OPT-NIR} \\
   & {\rm mean} & {\rm std} & {\rm mean} & {\rm std} & {\rm mean} & {\rm mean} \\
\noalign{\smallskip}
\hline
\noalign{\smallskip}
 0.21 & -0.17 \pm 0.04 & 0.35 \pm 0.07 & ..... & ..... & 2.4 \pm 0.1 & ..... \\
 0.31 & ..... & ..... & -0.40 \pm 0.14 & 0.6 \pm 0.1 & ..... & 3.8 \pm 0.2 \\
 0.64 & -0.33 \pm 0.17 & 0.80 \pm 0.14 & -0.38 \pm 0.15 & 0.56 \pm 0.15 & 2.6\pm 0.2& 3.6\pm0.2\\
\noalign{\smallskip}
\hline
\end{array}
$$
\end{table}

\subsection{Distributions of colour gradients}
\begin{figure}
\begin{center}
\resizebox{4cm}{9cm}{\includegraphics{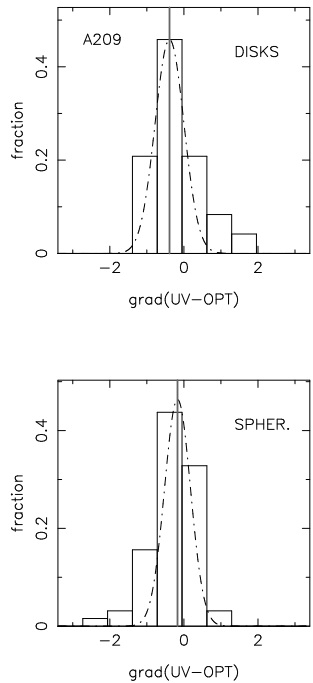}}
\end{center}
\caption[]{ Distribution of UV - OPT colour gradients for \a209. Upper
and lower panels refer to disks and spheroids, respectively.
\label{021CGRAD}
}
\end{figure}

\begin{figure}
\begin{center}
\resizebox{4cm}{9cm}{\includegraphics{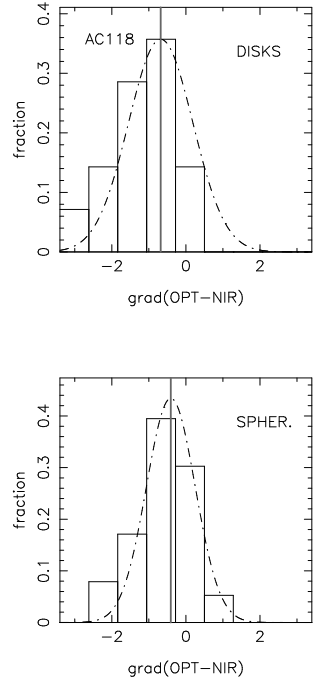}}
\end{center}
\caption[]{ The same of Fig.~\ref{021CGRAD} for the OPT -- NIR colour
gradients of galaxies in \ac118.
\label{031CGRAD}
}
\end{figure}

\begin{figure}
\begin{center}
\resizebox{9cm}{9cm}{\includegraphics{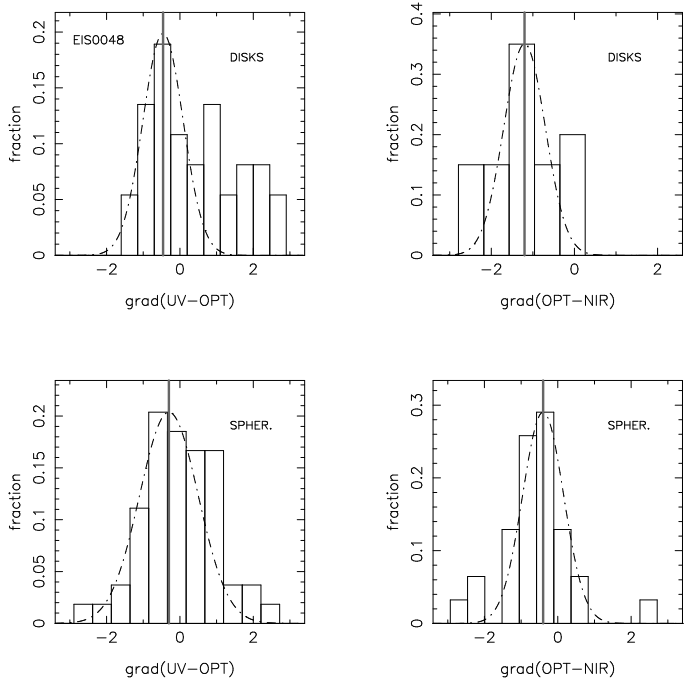}}
\end{center}
\caption[]{ The same of Fig.~\ref{021CGRAD} for the UV-OPT and OPT-NIR colour gradients of galaxies in \cl0048.
\label{064CGRAD}
}
\end{figure}

Colour gradients for \a209, \ac118 \, and \cl0048 \, are shown in
Fig.~\ref{021CGRAD},~\ref{031CGRAD} and~\ref{064CGRAD}, where upper
and lower panels denote disks and spheroids.  The bi-weight statistics
was applied in order to characterize each distribution by the relative
mean and standard deviation, whose values are summarized in
Tables~\ref{CGRAD_DISK} and~\ref{CGRAD_SPHER} for disks and spheroids,
respectively. In the same tables, we also report the mean values of
the UV-OPT and OPT-NIR central colours ($\mu_2 - \mu_1$ at
$\mathrm{r=r_i}$).  The relative distributions and the standard
deviations are not shown for brevity, since they are not relevant for
the following analysis. The UV-OPT and OPT-NIR central colours will be
used in Sect.~\ref{SECECGRAD} to discuss the origin of colour
gradients. We note that the $\mathrm{UV-OPT}$ colour indicates
$\mathrm{B-R}$ and $\mathrm{V-I}$ observed colours for \a209 \, and
\cl0048, respectively, while the $\mathrm{OPT-NIR}$ colour denotes
$\mathrm{R-K}$ and $\mathrm{I-K}$ for \ac118 \, and \cl0048,
respectively. As mentioned in Sect.~2, the values of
$\mathrm{grad(UV-OPT)}$ and $\mathrm{grad(OPT-NIR)}$ correspond
approximately to $\mathrm{grad(U-V)}$ and $\mathrm{grad(V-H)}$
restframe, respectively.

\subsubsection{Disks.}
\label{CGDISK}
The mean value of the $\mathrm{UV}-\mathrm{OPT}$ colour gradients
(hereafter ${\rm grad(UV-OPT)}$) amounts to $\mathrm{ \sim -0.4 }$ and
is fully consistent between \a209 \, and \cl0048. This is due to the
fact that the difference between $\mathrm{OPT}$ and $\mathrm{UV}$
structural parameters does not change significantly with redshift
(Sect.~\ref{PARDISK}).  We note, however, that the distribution of
\cl0048 \, is more disperse and asymmetric with respect to that of
\a209, with a significant fraction of galaxies ($\sim 40 \%$) having
${\rm grad(UV-OPT) > 0}$.  This is a consequence of what found in
Sect.~\ref{PARDISK}: some disks at $\mathrm{ z = 0.64 }$ have a light
profile more peaked toward the center in the UV with respect to the
optical, and show, therefore, positive values of ${\rm grad(UV-OPT)}$.

The most striking difference in $\mathrm{OPT}$ -- $\mathrm{NIR}$
colour gradients (${\rm grad(OPT-NIR)}$) between \ac118 \, and \cl0048
is the mean value of the two distributions. As shown in
Table~\ref{CGRAD_DISK}, the mean value of ${\rm grad(OPT-NIR)}$ is
$\sim -0.68$ at $\mathrm{ z=0.31 }$ and $\sim -1.2$ at $\mathrm{
z=0.64 }$, implying that there is a sharp evolution in the OPT -- NIR
colour gradients from $\mathrm{z=0.31}$ to $\mathrm{z = 0.64}$: the
NIR light of disks become much more peaked in the center at higher
redshift (cfr. Sect.~\ref{PARDISK}).

In order to account for field contamination\footnote{We point out
that, since galaxies were selected by the photometric redshifts, field
contaminants in the present sample are expected to lie approximately
at the same redshift of cluster galaxies. } in the previous results,
we proceeded as follows.  As discussed in Sect.~2 of Merluzzi et
al.~(\cite{MLM03}), the effect of contaminants for \ac118 is expected
to be negligible, while for \cl0048 the fraction of field contaminants
can be estimated as described in LMI03. By considering the colour
distribution of disks at $\mathrm{z=0.64}$, we estimated that $\sim 6$
(out of 20) galaxies in the OPT+NIR sample are expected to be field
galaxies, and applied a statistically subtraction procedure to correct
the distributions of colour gradients. The mean value of ${\rm
grad(OPT-NIR)}$ corrected for field contamination amounts to
$-1.35\pm0.07$, which is in agreement with that given in
Table~\ref{CGRAD_DISK}. The same result holds for UV-OPT colour
gradients.

\subsubsection{Spheroids.}
\label{CGSPHER}
By looking at Table~\ref{CGRAD_SPHER}, we see that the mean value of \,
${\rm grad(UV-OPT)}$ is $\sim -0.17$ at $\mathrm{ z=0.21 }$ and $\sim
-0.33$ at $\mathrm{ z=64 }$. While this difference is not particularly
significant ($< 1 \sigma $), the difference in the scatter of the two
distributions is much more evident, as it is clearly seen by comparing
Fig.~\ref{021CGRAD} (lower panel) and Fig.~\ref{064CGRAD} (lower --
left panel). The standard deviation of ${\rm grad(UV-OPT)}$ for
\cl0048 is more than double with respect to \a209. This is due to the
fact that, as discussed in Sect.~\ref{PARSPHER}, the dispersions of the
$\Delta (\log r_e)$ and $\Delta (\log n)$ distributions increase
significantly from $z=0.21$ to $z=0.64$.  

The mean values of ${\rm grad(OPT-NIR)}$ are $\sim - 0.51$ for \ac118
\, and $\sim -0.38$ for \cl0048, and are consistent within the
relative uncertainties. We conclude, therefore, that no strong
evolution in the OPT -- NIR colour gradients is found between
$\mathrm{z=0.31}$ and $\mathrm{z=0.64}$ for the population of
spheroids.

\section{Evolution of colour gradients}
\label{SECECGRAD}
To investigate the origins of the observed gradients, we compare our
results (colour gradients and central colours) with the predictions of
population synthesis models.

\subsection{Models.}
Colour gradients are modelled by using the GISSEL00 synthesis code
(Bruzual~\& Charlot \cite{BrC93}). Galaxies are described by an inner
stellar population, with age\footnote{Here and in the following age
refers to z=0.} $\mathrm{ T_i }$ and metallicity $\mathrm{ Z_i }$, and
by an outer stellar population, with age $\mathrm{ T_o }$ and
metallicity $\mathrm{ Z_o }$. Each of them is characterized by an
exponential star formation rate $\mathrm{e^{-t/\tau}}$. For spheroids,
we use $\mathrm{\tau = 1~Gyr}$, which gives a suitable description of
the evolution of their integrated colours (e.g. Merluzzi et
al.~\cite{MLM03}), while for the disks we construct models with both
$\mathrm{\tau=1~Gyr}$ and $\mathrm{ \tau = 3~Gyr}$, in order to
account for a star formation rate that lasts longer in time. For
disks, we also consider the effect of dust absorption, by describing
the inner and outer populations with colour excesses
$\mathrm{E(B-V)_i}$ and $\mathrm{E(B-V)_o}$, respectively. The
absorption in UV-OPT and OPT-NIR colours is computed by using the
extinction law from Seaton~(\cite{Sea79}).  For \a209, \ac118 \, and
\cl0048, the colours of the inner and outer populations are obtained
by using the instrumental response curves\footnote{The response curves
were retrieved from the ESO Exposure Time Calculators at
http://www.eso.org/observing/etc/.}, while colour gradients at
$\mathrm{z\sim0}$ are computed by using the U-, V-, and K-band filter
curves from Buser~(\cite{BUS78}), Azusienis \&
Straizys~(\cite{AzS69}), and Bessell \& Brett~(\cite{BeB88}),
respectively.

Each gradient model is characterized by a set of parameters
$\mathrm{P_k}$ ($\mathrm{Z_i, T_i, ...}$), whose value is derived from
the observations. To this aim, we minimize the chi-square: 
\begin{equation} 
\mathrm{ 
 \chi^2(P_k)=\sum[(E_k-O_k)^2/\sigma_k^2 ], \label{EQCHI}
}
\end{equation}
where $\mathrm{O_k}$ and $\mathrm{\sigma_k}$ are the observed
quantities (central colours and colour gradients) and the relative
uncertainties at the different redshifts, while $\mathrm{E_k}$ are the
model predictions for $\mathrm{O_k}$.

\subsubsection{Spheroids.}
We consider three kinds of gradient models: metallicity (Z) models,
age (T) models, and metallicity+age (TZ) models.  The relevant
information on the models are summarized in Table~\ref{CG_MOD_SPHER}.
\begin{table}[ht]
\caption[]{Parameters of colour gradient models of spheroids.  Column
1: model. Columns 2, 3: inner and outer metallicities.  Columns 4, 5:
inner and outer ages. Notes: (a) constrained by both central colours
and colour gradients, (b) the parameters are chosen in order to
describe extremely negative colour gradients, (c) constrained by
grad(UV-OPT) at $\mathrm{z=0}$, (d) the parameters are chosen in order
to describe positive colour gradients.
}
\label{CG_MOD_SPHER}
$$ 
\begin{array}{lcccc}
\hline
\noalign{\smallskip}
  \mathrm{MOD} & \mathrm{Z_i} & \mathrm{Z_o} & \mathrm{T_i} & \mathrm{T_o} \\
   (1) & (2) & (3) & (4) & (5) \\
\noalign{\smallskip}
\hline
\noalign{\smallskip}
\mathrm{Z1} & \mathrm{3/2Z_{\odot}}^{(a)} & \mathrm{0.6Z_{i}}^{(a)}  & \mathrm{13~Gyr} & \mathrm{13~Gyr} \\
\mathrm{Z2}^{(b)} & \mathrm{3Z_{\odot}}   & \mathrm{1/20Z_{i}} & \mathrm{13~Gyr} & \mathrm{13~Gyr} \\
\mathrm{T1} & \mathrm{Z_{\odot}} & \mathrm{Z_{\odot}} & \mathrm{13~Gyr} & \mathrm{0.62T_i}^{(c)}  \\
\mathrm{T2}^{(d)} & \mathrm{3Z_{\odot}} & \mathrm{3Z_{\odot}} & \mathrm{0.8T_i}  & \mathrm{13~Gyr}  \\ 
\mathrm{TZ} & \mathrm{3/2Z_{\odot}}^{(a)} & \mathrm{1/2Z_{\odot}}^{(a)} & \mathrm{13~Gyr} & \mathrm{0.85T_i}^{(a)}\\
\noalign{\smallskip}
\hline
\end{array}
$$
\end{table}
For models (Z), we assume that the inner and outer populations are
coeval, with $\mathrm{T_i = T_o = 13~Gyr}$, while the values of
$\mathrm{ Z_i }$ and $\mathrm{ Z_o }$ are changed.  We consider two
cases: model $\mathrm{ (Z1) }$, whose inner and outer metallicities
are constrained by the observations, and model $\mathrm{(Z2)}$, which
describes a strong metallicity gradient, with $\mathrm{Z_i =
3Z_{\odot}}$ and $\mathrm{Z_o = 1/20Z_i}$. We note that inverse
metallicity models ($\mathrm{Z_o > Z_i}$) are not considered here
since they are not consistent with the data (see
Sect.~\ref{MOD_OBS}). For models (T), we assume that the inner and the
outer stellar populations have the same metallicity, $\mathrm{Z_i =
Z_o}$, while $\mathrm{T_o}$ and $\mathrm{T_i}$ are changed.  We
consider two cases: model $\mathrm{ (T1) }$ has $\mathrm{T_i =
13~Gyr}$, $\mathrm{Z_i = Z_\odot}$ and $\mathrm{T_o < T_i}$, while
model $\mathrm{ (T2) }$ has $\mathrm{Z_i = 3Z_\odot}$, $\mathrm{T_o =
13~Gyr}$ and a younger population in the inner region. The value of
$\mathrm{T_o}$ for model $\mathrm{ (T1) }$ is derived on the basis of
the observations while for model $\mathrm{ (T2) }$ we chose
$\mathrm{T_i = 0.8 \cdot T_o}$.  We note that model $\mathrm{ (T2) }$
describes a galaxy in which the stars redden toward the outskirts, and
is considered in order to investigate the origin of positive colour
gradients.  Finally, in model (TZ) the inner population is old
($\mathrm{T_i = 13~Gyr}$), while $\mathrm{Z_i}$, $\mathrm{Z_o}$ and
$\mathrm{T_o}$ are obtained from the observations.

\subsubsection{Disks.}
The relevant information on the gradient models are summarized in
Table~\ref{CG_MOD_DISK}.  We consider the following models: metallicity
(Z) and age (T) models, a `protracted' age ($\mathrm{ \widetilde{T}1
}$) model, a protracted age+dust ($\mathrm{\widetilde{T}D}$) model,
and a `changing' dust ($\mathrm{D}$) model.  For (Z) and (T) models,
we consider only the models (Z1) and (T1) introduced for the
spheroids, since (Z2) and (T2) are not relevant for the present
discussion.  Model ($\mathrm{ \widetilde{T}1 }$) is similar to (T1)
but is characterized by a more protracted star formation 
($\mathrm{\tau=3~Gyr}$) in both the inner and outer regions. The same
value of $\tau$ is used for models ($\mathrm{\widetilde{T}D}$) and
($\mathrm{D}$). Model ($\mathrm{\widetilde{T}D}$) differs from
($\mathrm{ \widetilde{T}1 }$) for the presence of dust in the inner
and outer regions. The values of $\mathrm{E(B-V)_i}$ and
$\mathrm{E(B-V)_o}$ are assumed to be equal and are derived from the
observations. Finally, model ($\mathrm{D}$) has
$\mathrm{Z_i=Z_o=Z_\odot}$, $\mathrm{T_i=T_o=13~Gyr}$, and
$\mathrm{E(B-V)_o=0}$. The variation of colour gradients with redshift
is described by changing the amount of dust absorption in the inner
region between $z \le 0.31$ and $\mathrm{z=0.64}$.
\begin{table}[ht]
\caption[]{Parameters of colour gradient models of disks.  Column 1:
model. Columns 2, 3: inner and outer metallicities.  Columns 4, 5:
inner and outer ages. Columns 6, 7: inner and outer colour
excesses. Column 8: time scale of the star formation rate.  Notes: (a)
constrained by both central colours and $\mathrm{grad(UV-OPT)}$ at
$\mathrm{z=0.21}$ and $\mathrm{z=0.64}$, (b) constrained by the colour
gradients, (c) constrained by both central colours and UV-OPT, OPT-NIR
colour gradients, (d) the first value of $\mathrm{E(B-V)_i}$ refers to
$\mathrm{z=0.21}$ and $\mathrm{z=0.31}$, while the second to
$\mathrm{z=0.64}$.  }
\label{CG_MOD_DISK}
$$ 
\begin{array}{lccccccc}
\hline
\noalign{\smallskip}
  \mathrm{MOD} & \mathrm{Z_i} & \mathrm{Z_o} & \mathrm{T_i} & \mathrm{T_o} & \mathrm{E(B-V)_{i}} & \mathrm{E(B-V)_{o}} & \mathrm{\tau} \\
  (1) & (2) & (3) & (4) & (5) & (6) & (7) & (8) \\
\noalign{\smallskip}
\hline
\noalign{\smallskip}
\mathrm{Z1} &  \mathrm{3Z_{\odot}}^{(a)} & \mathrm{0.8Z_{\odot}}^{(a)} & \mathrm{13~Gyr} & \mathrm{13~Gyr} & 0. & 0. & \mathrm{1~Gyr} \\
\mathrm{T1} & \mathrm{0.44Z_\odot}^{(b)} & \mathrm{0.44Z_\odot}^{(b)} & \mathrm{13~Gyr} & \mathrm{0.55T_i}^{(b)} & 0. & 0. & \mathrm{1~Gyr} \\
\mathrm{\widetilde{T}1} & \mathrm{0.44Z_\odot}^{(b)} & \mathrm{0.44Z_\odot}^{(b)} & \mathrm{13~Gyr} & \mathrm{0.55T_i}^{(b)} & 0. & 0. & \mathrm{3~Gyr} \\
\mathrm{\widetilde{T}D} & \mathrm{0.44Z_\odot}^{(c)} & \mathrm{0.44Z_\odot}^{(c)} & \mathrm{13~Gyr} & \mathrm{0.55T_i}^{(c)} & 0.18^{(c)} & 0.18^{(c)} & \mathrm{3~Gyr} \\
\mathrm{\mathrm{D}}^{(d)} & \mathrm{Z_\odot} & \mathrm{Z_\odot} & \mathrm{13~Gyr} & \mathrm{13~Gyr} & 0.15\!,\!0.30^{(d)} & 0. & \mathrm{3~Gyr} \\
\noalign{\smallskip}
\hline
\end{array}
$$
\end{table}

\subsection{Comparison with observations.}
\label{MOD_OBS}
\begin{figure}[ht]
\begin{center}
\resizebox{12cm}{12cm}{\includegraphics{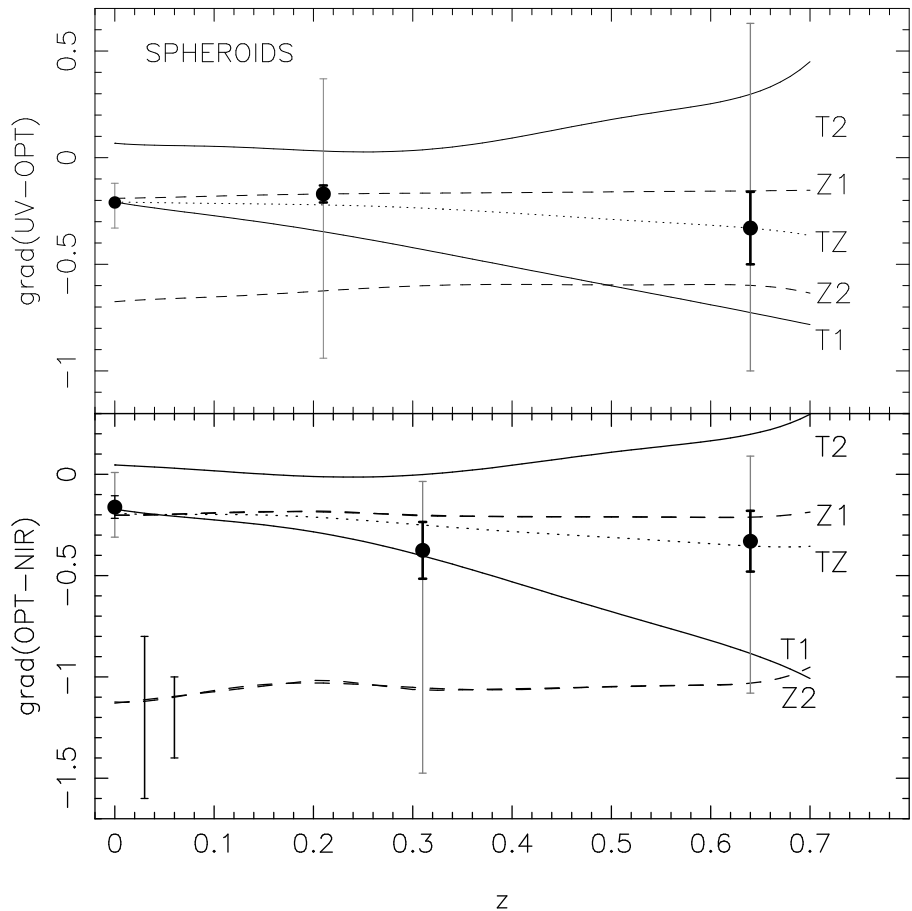}}
\end{center}
\caption[]{ $\mathrm{UV}-\mathrm{OPT}$ and $\mathrm{OPT}-\mathrm{NIR}$
colour gradients of spheroids as a function of redshift. The black and
the grey error bars denote, respectively, the uncertainties on the
mean value of the colour gradient and the $32\%$ percentile interval
of the distribution of colour gradients.  Metallicity and age models
are represented by the dashed and the continuous curves. The typical
uncertainties on individual colour gradients are shown by the error
bars in the lower -- left of the bottom panel for \ac118 and \cl0048
(left and right). Typical errors for A\,209 are intermediate.

\label{E_CGRAD_S}
}
\end{figure}

\begin{figure}[ht]
\begin{center}
\resizebox{12cm}{12cm}{\includegraphics{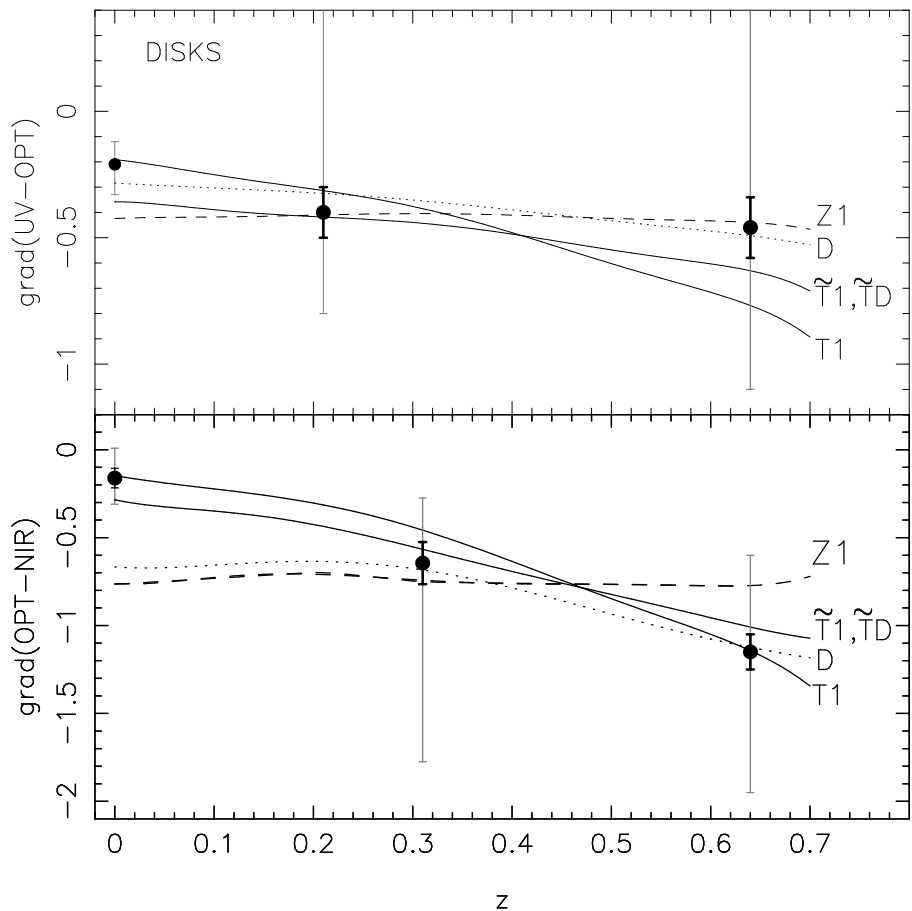}}
\end{center}
\caption[]{ Same of Fig.~\ref{E_CGRAD_S} for the disks.
\label{E_CGRAD_D}
}
\end{figure}

The observed colour gradients are compared with the models in
Figs.~\ref{E_CGRAD_S} and ~\ref{E_CGRAD_D} for spheroids and disks,
respectively. The curves in the figures were obtained by interpolating
the values of the colour gradient models at $z=0, 0.21, 0.31$ and
$0.64$.

To further constrain the models, we also considered measurements of
colour gradients for nearby galaxies. For the spheroids, we use the
$\mathrm{UV}$ -- $\mathrm{OPT}$ values for a sample of 39 nearby
ellipticals by PDI90, and the $\mathrm{OPT}$ -- $\mathrm{NIR}$ colour
gradients obtained by PVJ90 for 19 elliptical galaxies. For the disks,
an homogeneous sample for which colour gradient measurements are
available at $\mathrm{z \sim 0}$ both in UV -- OPT and in OPT -- NIR
is not available, and therefore we consider, as reference, the values
relative to the spheroids. In the figures, the black symbols denote
the mean values of the colour gradients and the relative uncertainties
($1 \sigma$ standard intervals), while the grey bars, which mark the
$32 \%$ percentile intervals of the gradient distributions, describe
the range of values of colour gradients at each redshift.

The calibration of the models was performed by minimizing the
`distance' (see Eq.~(\ref{EQCHI})) of the model predictions from the
observed quantities (central colours and/or colour gradients), as
summarized in Tables~\ref{CG_MOD_SPHER} and~\ref{CG_MOD_DISK}.

\subsubsection{Spheroids.} 
By looking at Fig.~\ref{E_CGRAD_S}, we begin to note that the size
of the grey bars at $z \sim 0$ are much smaller than those at higher
redshift. Considering the typical uncertainty on the estimated colour
gradients, we see that the difference is not due to the uncertainty on
the estimate of structural parameters. The simplest explanation
resides in the different selection criteria of our samples and those
of PDI90 and PVJ90: the samples at $z \sim 0$ consist, in fact, of
morphologically selected elliptical galaxies, that, in many regards,
are known to have very homogeneous properties (see e.g. van Dokkum et
al. \cite{vDF98}). To discuss the dispersion in the values of colour
gradients, we will consider, therefore, only the estimates at
$\mathrm{ z > 0 }$.

Fig.~\ref{E_CGRAD_S} shows that model $\mathrm{Z1}$ gives a good
description of all the observed colour gradients. Moreover this model
produces the following central colours:
$\mathrm{(UV-OPT)}_{0.21}\sim2.65$, $\mathrm{(UV-OPT)}_{0.64}=2.58$,
$\mathrm{(OPT-NIR)}_{0.31}\sim3.7$, $\mathrm{(OPT-NIR)}_{0.64}=3.2$.
These values are consistent (within at most $2\sigma$) with the
observations (see Table~\ref{CGRAD_SPHER}).  We also note that model
$\mathrm{Z1}$ produces very small $\mathrm{OPT}$ -- $\mathrm{OPT}$
colour gradients in the considered redshift range
(e.g. $\mathrm{grad(R-I)_{0.31}=-0.04}$), and is therefore in
agreement with optical -- optical estimates obtained by other studies
for cluster galaxies at intermediate redshift (e.g. Saglia et
al.~\cite{SMG00}, Tamura \& Ohta~\cite{TaO00}).  The age gradient
model (T1) is consistent with the values of $\mathrm{ grad(OPT-NIR)}$
at $\mathrm{ z \sim 0 }$ and at $\mathrm{ z \sim 0.31 }$, as discussed
in LBM02. However the other values of colour gradients exclude
definitively this model, confirming that metallicity is the primary
cause driving the colour gradients.  For what concerns model
$\mathrm{TZ}$, we see that the combined effect of age and metallicity
can explain to some extent both the colour gradients and the central
colours at all redshifts. This model produces, in fact, the same
central colours of model $\mathrm{Z1}$ and is therefore in agreement
with the values in Table~\ref{CGRAD_SPHER}. On the other hand, the very
rapid decline of $\mathrm{grad(UV-OPT)}$ and $\mathrm{grad(OPT-NIR)}$
produced by age gradients (model $\mathrm{ T1 }$) as a function of
redshift, strongly constrain the maximum allowed age variation,
$\mathrm{\Delta( T)_{max}=(T_i - T_o)_{max}}$, from the galaxy center
to the outskirts. In order to estimate $\mathrm{\Delta(T)_{max}}$, we
shifted the values of the colour gradients and of the central colours
according to the relative uncertainties by using normal random
deviates. For each iteration, we re-derived the values of
$\mathrm{Z_o}$ and $\mathrm{T_o}$ of model $\mathrm{(TZ)}$. We found
that the values of $\mathrm{Z_o}$ and $\mathrm{T_o}$ change according
to the relation $\mathrm{ \delta( \log T_o) \sim -2/3 \delta( \log
Z_o) }$, which describes the well known age--metallicity degeneracy
(Worthey et al.~\cite{WTF96}), and that $\sim 90 \%$ of the
$\mathrm{T_o}$ values are greater than $\mathrm{\sim 10~Gyr}$,
constraining $\mathrm{\Delta(T)}$ to be smaller than $\sim 23\%$ of
$\mathrm{T_i}$.

For what concerns the grey bars, we see that strong metallicity
gradients (model $\mathrm{Z2}$) are able to describe the tail toward
negative values of the colour gradient distributions. The $\mathrm{
OPT }$ -- $\mathrm{ NIR }$ diagram excludes, however, metallicity
gradients stronger than that of model $\mathrm{Z2}$, setting the
constraint $\mathrm{Z_o \widetilde{>} 0.15 \cdot Z_\odot} $.  Positive
values of $\mathrm{grad(UV - OPT)}$ can be described by age models
with $\mathrm{T_i < T_o}$, like model $\mathrm{T2}$. Inverse
metallicity models ($\mathrm{Z_i < Z_o}$) give positive values of
$\mathrm{grad(UV - OPT)}$ but too high values of $\mathrm{grad(OPT -
NIR)}$, and cannot fit, therefore, both gradients simultaneously.  The
$\mathrm{ OPT }$ -- $\mathrm{ NIR }$ diagram excludes colour gradients
significantly higher than those of model $\mathrm{T2}$, setting the
constraint $\mathrm{ T_i > 0.8 \cdot T_o }$.

\subsubsection{Disks.} 
By looking at Fig.~\ref{E_CGRAD_D}, it is evident that metallicity
models fail to reproduce the variation of $\mathrm{grad( OPT - NIR )}$
with redshift. Moreover, metallicity gradients in local disks are
known to be small (e.g. Zaritsky et al.~\cite{ZKH94}) and therefore
they cannot account for the large colour gradients observed in the
present samples.

On the contrary, age models ($\mathrm{ T1 }$ and $\mathrm{
\widetilde{T}1 }$) give a better description of the data. In model
$\mathrm{ T1 }$, $\mathrm{grad( UV - OPT )}$ varies too rapidly with
redshift, deviating by $\sim3\sigma$ from the observed gradient at
$z=0.64$. Increasing the $\tau$ value, the variation of $\mathrm{grad(
UV - OPT )}$ becomes milder, and the agreement improves\footnote{The
UV-OPT colour is more sensitive with respect to OPT-NIR colours to the
presence of young stars. For high values of $\tau$, the inner and
outer stellar populations protract the formation of fresh stars with
redshift, and therefore the evolution of $\mathrm{grad(UV-OPT)}$
becomes smaller with respect to that of $\mathrm{grad(OPT-NIR)}$.}.
Model ($\mathrm{\widetilde{T}1}$) is able, in fact, to reproduce
within at most $1.3\sigma$ the values of $\mathrm{grad( UV - OPT )}$
and $\mathrm{grad( OPT - NIR )}$ at any redshift, and gives colour
gradients at $\mathrm{z \sim 0}$ which are only slightly bluer than
those of the spheroids.  Although age models reproduce the observed
colour gradients, they fail to describe the red central colours of
disks (Table~\ref{CGRAD_DISK}), which are not significantly different
from those of the spheroids (Table~\ref{CGRAD_SPHER}).  Model
$\mathrm{\widetilde{T}1}$ gives $\mathrm{(UV-OPT)}_{0.21}\sim2.0$,
$\mathrm{(UV-OPT)}_{0.64}\sim1.9$, $\mathrm{(OPT-NIR)}_{0.31}\sim3.0$,
$\mathrm{(OPT-NIR)}_{0.64}=2.5$, which are not consistent with the
observations. We also note that, by increasing the inner and outer
metallicities, the central colours are better reproduced, but the
description of colour gradients worsens\footnote{In fact,
$\mathrm{grad(UV-OPT)}$ is more sensitive to the metallicity with
respect to $\mathrm{grad(OPT-NIR)}$, in the sense that increasing
$\mathrm{Z_i}$ and $\mathrm{Z_o}$, the UV-OPT gradients decrease more
rapidly than $\mathrm{grad(OPT-NIR)}$.}.  By fitting $\mathrm{T_i,
T_o, Z_i, Z_o}$ to both the colour gradients and the central colours,
we found that these parameters are not able by themselves to describe
all the observed quantities at better than $2.5\sigma$. We conclude,
therefore, that stellar population parameters cannot explain alone our
data.

The possible solution resides in dust absorption, which is known to
produce strong colour gradients in nearby late-type spirals (Peletier
et al.~\cite{PVM95}). Model $\mathrm{\widetilde{T}D}$ gives the same
observed gradients of model $\mathrm{\widetilde{T}1}$, and produces
the following central colours: $\mathrm{(UV-OPT)}_{0.21}\sim2.4$,
$\mathrm{(UV-OPT)}_{0.64}\sim2.3$, $\mathrm{(OPT-NIR)}_{0.31}\sim3.9$,
$\mathrm{(OPT-NIR)}_{0.64}=3.3$, in good agreement ($\le1.5\sigma$)
with the values in Table~\ref{CGRAD_DISK}. We point out that the
present data can also be explained by models in which both age and
dust content vary toward the periphery.  In order to investigate this
point, we considered $\mathrm{T_o}$ and $\mathrm{E(B-V)_o}$ in model
$\mathrm{\widetilde{T}D}$ as free parameters and fitted them to the
observed quantities. The procedure was iterated by shifting the
central colours and the colour gradients according to the
corresponding uncertainties. We found that the change of
$\mathrm{T_o}$ and $\mathrm{E(B-V)_o}$ are correlated, in the sense
that $\mathrm{\delta(\log T_o)} \sim -1/3 \mathrm{\delta(\log
E(B-V)_o)}$, and that the variation of $\mathrm{E(B-V)}$ and
$\mathrm{T}$ from the center to the periphery,
$\mathrm{\delta(E(B-V))}$ and $\mathrm{\delta(T)}$, are constrained in
the ranges $ 8\% \lesssim \mathrm{\delta(E(B-V))} \lesssim 17\% $ and
$45\% \lesssim \mathrm{\delta(T)} \lesssim 48 \%$ of the central
values\footnote{The extrema mark a $10\%$ percentile interval}.  By
increasing the inner and outer metallicities to
$\mathrm{Z_i=Z_o=3/2Z_\odot}$, the best fit values of
$\mathrm{E(B-V)_i}$ is $\sim0.09$, and the following constraints are
obtained: $ 30\% \lesssim \mathrm{\delta(E(B-V))} \lesssim 90\% $ and
$30\% \lesssim \mathrm{\delta(T)} \lesssim 45 \%$. By increasing
$\tau$ to $\mathrm{10~Gyr}$ (with $\mathrm{Z_i=Z_o=0.44Z_\odot}$), we
obtain $\mathrm{E(B-V)_i}=0.26$, $ 51\% \lesssim
\mathrm{\delta(E(B-V))} \lesssim 53\% $ and $40\% \lesssim
\mathrm{\delta(T)} \lesssim 45 \%$.  To summarize, in order to explain
central colours and colour gradients of disks, both age gradients and
dust absorption are needed. For a wide range of stellar population
parameters, the radial age variation is strongly constrained in the
range $[30,50]\%$ of the central value, while $\mathrm{E(B-V)_i}$ is
greater than $\sim0.1$.  On the other hand, the radial gradient of
dust absorption depends critically on the inner metallicity and on the
time scale of star formation rate.

\begin{figure}[ht]
\begin{center}
\resizebox{10cm}{10cm}{\includegraphics{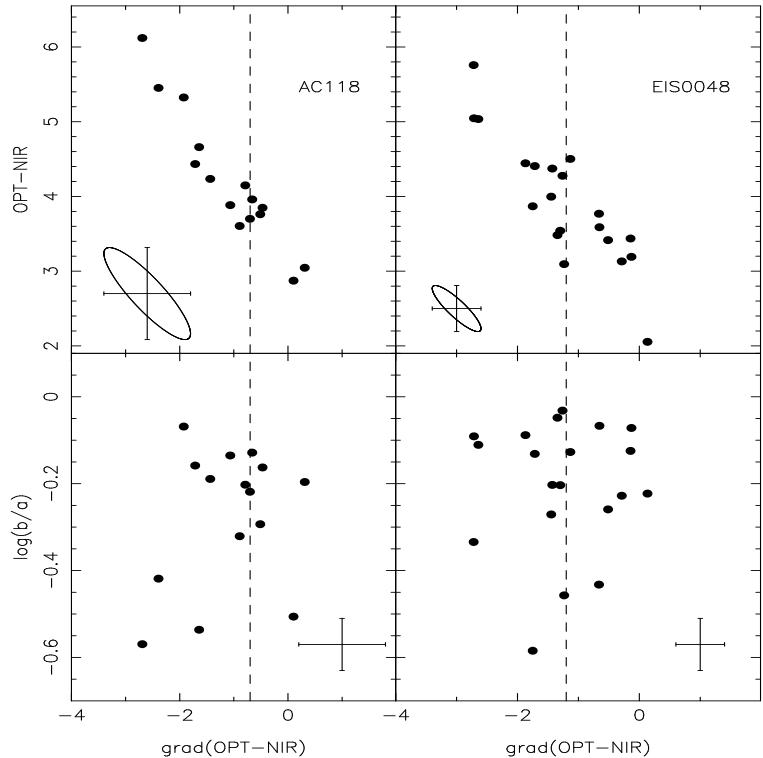}}
\end{center}
\caption[]{OPT-NIR central colours and logarithm of axis ratio
vs. OPT-NIR colour gradients for AC118 (left panels) and EIS0048 (right
panels). Dashed lines mark the mean values of ${\rm grad(OPT-NIR)}$.
The ellipses in the upper panels correspond to $1\sigma$ confidence
contours and show the correlation of the errors on OPT-NIR colours and
colour gradients.
\label{CGRAD_CC_BA}
}
\end{figure}

So far, we considered only models for which the dust gradient does not
change from one cluster to the other.  By relaxing this assumption,
the presence of age gradients is no more needed in order to explain
all the observed quantities. This is shown by model $\mathrm{D}$: a
difference of the dust gradient between $\mathrm{z\le0.31}$ and
$\mathrm{z=0.64}$ also fits the data, giving central colours
$\mathrm{(UV-OPT)}_{0.21}\sim2.4$, $\mathrm{(UV-OPT)}_{0.64}\sim2.5$,
$\mathrm{(OPT-NIR)}_{0.31}\sim3.9$, $\mathrm{(OPT-NIR)}_{0.64}=3.8$,
which are in good agreement (at $\le1.5\sigma$) with the values in
Table~\ref{CGRAD_DISK}.  As shown by Peletier et al.~\cite{PVM95},
useful tools to investigate the role of dust gradients in galaxies are
the diagrams showing colour gradients against central colours and/or
inclination. If dust is the primary effect driving the observed
gradients (model $\mathrm{D}$), we would expect that redder and more
inclined disks have also stronger OPT-NIR colour gradients for both
\ac118 and \cl0048. The diagrams are shown in
Fig.~\ref{CGRAD_CC_BA}. By analyzing the figure, it is important to
take into account that the uncertainties on central colours and colour
gradients are correlated.  At least for \ac118, this correlation fully
explains the tight relation between $\mathrm{grad(OPT-NIR)}$ and the
$\mathrm{OPT-NIR}$ colours, while for \cl0048 some correlation could
exist. On the other hand, the diagrams with inclination
vs. $\mathrm{grad(OPT-NIR)}$ do not show any correlation at all
(cfr. Fig.~1 of Peletier et al.~\cite{PVM95}).  These facts seem to
suggest that dust is not the primary cause of the observed gradients,
and that models $\mathrm{\widetilde{T}D}$ are a more suitable
explanations of the present data.

Finally, some considerations are needed about the ranges of values of
$\mathrm{grad( UV - OPT )}$ and $\mathrm{grad( OPT - NIR )}$.  We see
that age gradients stronger than $\mathrm{T1}$ can successfully
describe the tails toward negative values of these distributions, with
the constrain $\mathrm{T_o > 0.5 \cdot T_i}$ set by $\mathrm{grad( UV
- OPT )}$ at $\mathrm{z=0.64}$. For what concerns the positive values
of $\mathrm{grad( UV - OPT )}$, which correspond to negative
$\mathrm{OPT-NIR}$ gradients, they seem not to be explainable with the
models considered above.

\section{Summary and conclusions.}
\label{SECCONC}
We have studied the waveband dependence of the shape of the light
distribution in cluster galaxies from $\mathrm{z\sim0.2}$ to
$\mathrm{z\sim0.64}$, by using a large wavelength baseline covering UV
to NIR restframe.  New structural parameters have been derived for the
galaxy clusters \a209 \, at $z=0.21$, in UV and OPT restframe (B and R
bands), and for \cl0048, at $\mathrm{z = 0.64}$, in UV, OPT and NIR
restframe (V, I and K bands). Previous determinations of OPT and NIR
(R and K bands) structural parameters for the cluster \ac118 at
$\mathrm{z=0.31}$ have also been used. The present data constitute the
first large ($\mathrm{N \sim 270}$) sample of cluster galaxies with
surface photometry from UV to NIR at intermediate redshifts.  The
analysis has been performed in terms of both structural parameters and
colour gradients/central colours of galaxies, for the populations of
disks and spheroids.
\subsection{Spheroids.}
On average, the light profile of galaxies is more concentrated at
longer wavelengths: effective radii decrease from UV to NIR, while
Sersic indices increase. The values of $\mathrm{\Delta (\log r_e)}$
and $\mathrm{\Delta (\log n)} $ seem not to evolve significantly with
redshift, although some differences (at $\sim2\sigma$) among the
cluster could exist. By averaging the results for \a209, \ac118 \, and
\cl0048, we obtain $\mathrm{ r_e^{UV}/r_e^{OPT} = 1.2 \pm 0.05}$,
$\mathrm{ n^{OPT}/n^{UV} = 1.0 \pm 0.1 }$, and
$\mathrm{r_e^{OPT}/r_e^{NIR} = 1.26 \pm 0.06}$, $\mathrm{
n^{OPT}/n^{NIR} = 0.88 \pm 0.03 }$.  The value of
$\mathrm{r_e^{OPT}/r_e^{NIR}}$ is fully consistent with that obtained
for nearby early-type galaxies by Pahre et al.~\cite{PCD98}, who found
$\mathrm{r_e^{V}/r_e^{K}} \sim 1.2$. The ratio of UV--OPT effective
radii at $\mathrm{z\sim0}$ can be estimated from the U- and R-band
data of galaxies in rich nearby clusters by J\o rgensen et
al.~\cite{JFK95}, and amounts to $\mathrm{r_e^{U}/r_e^{R}} \sim 1.17$,
which is in full agreement with our results. The bulk of spheroids
shows negative colour gradients, both in UV -- OPT and in OPT -- NIR,
and is therefore characterized by stellar populations redder in the
center than in the periphery.  The mean values of colour gradients are
$\mathrm{grad(UV-OPT)=-0.18\pm0.04}$ and
$\mathrm{grad(OPT-NIR)=-0.4\pm0.1}$, and are consistent with the
values obtained by PDI90, $\mathrm{grad(U-R)=-0.21\pm0.02}$, by Idiart
et al.~(\cite{IPM03}), $\mathrm{grad(U-V)=-0.23\pm0.07}$, and by
PVJ90, $\mathrm{grad(V-K)=-0.16\pm0.06}$, for nearby
early-types. Moreover the UV-OPT colour gradient is fully consistent
with that estimated by Tamura \& Ohta~(\cite{TaO00}) for cluster
early-types at $\mathrm{z\sim0.4}$,
$\mathrm{grad(B_{450}-I_{814})}=-0.23\pm0.05$.  We conclude,
therefore, that the colour radial variation in spheroids does not
change significantly at least out to $\mathrm{z=0.64}$. The comparison
of colour gradients and central colours with gradient models of
stellar population parameters shows that metallicity is the primary
cause of the colour variation in galaxies, in agreement with previous
optical studies at intermediate redshifts. The data are well fitted by
models with super-solar inner metallicity ($\mathrm{Z_i=3/2Z_\odot}$),
and with a pure metallicity gradient of about $\mathrm{-0.2 dex}$ per
decade of radius, in agreement with what found by Saglia et
al.~(\cite{SMG00}) and Idiart et al.~(\cite{IPM03}).  Taking into
account age--metallicity degeneracies, however, mild age gradients are
also consistent with our results, the age of stellar populations
(referred to $\mathrm{z=0}$) being constrained to vary by less than
$25\%$ from the center to the outskirts.

\subsection{Disks.}
The light profiles of galaxies become more peaked in the center from
UV to NIR. For what concerns structural parameters, we find that
effective radii decrease at longer wavelengths, while Sersic indices
become greater. The ratios of UV--OPT parameters and OPT--NIR
effective radii do not show significant evolution with redshift. The
mean values of $\mathrm{r_e^{UV}/r_e^{OPT}}$,
$\mathrm{n^{UV}/n^{OPT}}$ and $\mathrm{r_e^{OPT}/r_e^{NIR}}$ amount to
$1.1\pm0.035$, $0.84\pm0.03$ and $1.25\pm0.04$, respectively.  The
most sharp result, however, is the decrease in the ratio of OPT to NIR
Sersic indices from $\mathrm{z=0.31}$ to $\mathrm{z=0.64}$: the value
of $\mathrm{n^{OPT}/n^{NIR}}$ varies from $\sim0.7$ at
$\mathrm{z=0.31}$ down to $\sim0.45$ at $\mathrm{z=0.64}$, implying
that the OPT--NIR structure of disks becomes significantly more
concentrated at higher redshift. When analyzed in terms of colour
gradients, the previous results entail a presence of UV--OPT and
OPT--NIR gradients much stronger with respect to those of spheroids.
This is similar to what found in studies of colour gradients for
nearby galaxies (e.g. Peletier et al.~\cite{PVM95}). We find that the
UV--OPT gradient does not vary significantly with redshift, and
amounts to $\sim -0.42$, while, on the contrary, the value of
$\mathrm{grad(OPT-NIR)}$ strongly decreases from $\sim-0.7$ at
$\mathrm{z=0.31}$ down to $\sim-1.2$ at $\mathrm{z=0.64}$. Moreover,
disks have redder colours in the center, comparable to those of
spheroids. We analyze these results by using simple gradient models
describing the effect of variations of stellar population parameters
as well as dust extinction throughout the galaxies. It turns out that
two different kinds of models fit the present data.
\begin{description}
\item[1.] Age gradient models with (i) a protracted star formation
rate both in the inner and in the outer regions, and with (ii) a
significant amount of dust extinction in the center. The property (i)
keeps the variation of $\mathrm{grad(UV-OPT)}$ mild and allows the
observed evolution of $\mathrm{grad(OPT-NIR)}$ to be reproduced, while
the property (ii) is needed in order to describe the red central
colours in disks. The better constrained quantities are the age
gradient, in the range $[30,50]\%$ of the central value, and the inner
extinction, $\mathrm{E(B-V)_i \widetilde{>} 0.1}$.  On the other hand,
the amount of dust in the outer region depends critically on the value
of other stellar population parameters (inner metallicity and time
scale of star formation). It is interesting to note that the age
gradient model predicts an $\mathrm{UV-OPT}$ colour gradients of $\sim
-0.35$ at $\mathrm{z\sim0}$, which is in very good agreement with that
measured by Gadotti \& dos Anjos~(\cite{GAD01}) for a sample of nearby
late-type spirals, $\mathrm{grad(UV - OPT)} \sim -0.33$ (see their
Table~2). The existence of age gradients in disk dominated galaxies is
in agreement with what found by de Jong~(\cite{deJ96}) for nearby
spirals, while the fact that dust absorption plays an important role
in disks is in agreement with the findings of Peletier et
al.~\cite{PVM95} for Sb and Sc galaxies at
$\mathrm{z\sim0}$. Moreover, age gradients are a robust prediction of
hierarchical models, in which disks are accreted by the condensation
of gas from the surrounding halos.
\item[2.] Pure extinction models, in which the gradient of
$\mathrm{E(B-V)}$ increases by a factor $\sim2$ for \cl0048 with
respect to the other clusters.
\end{description}
In order to discriminate between the two models, we analyze the
relations between colour gradients and central colours/axis ratios.
The present data indicate that galaxies with higher inclination do not
have stronger colour gradients, as one would expect if dust gradients
would be the main effect driving the observed gradients.  Although
some correlation between colour gradients and inclination could be
hidden by the uncertainties on the above quantities, we argue that age
gradient models could give a more suitable explanation of the present
data.

Since hierarchical merging scenarios predict strong differences in the
properties of field and cluster galaxies, it will be very interesting
to analyze the role that environment has on the internal colour
distribution of galaxies. Although we show by a statistical
subtraction procedure that our results are not affected from field
contamination, we cannot discriminate between the properties of field
and cluster galaxies. In order to address this point, spectroscopic
data will be needed.

\begin{acknowledgements}
We thank the ESO staff who effectively attended us during the
observation run at FORS2. We thank M. Capaccioli and G. Chincarini for
useful discussions. We are grateful to the referee, R. Peletier, for
his comments, which helped us to improve the manuscript.  Michele
Massarotti is partly supported by a MIUR-COFIN grant.
\end{acknowledgements}

\appendix
\section{Reduction of FORS2 I-band images.}
\begin{figure}
\begin{center}
\resizebox{11.7cm}{5.85cm}{\includegraphics{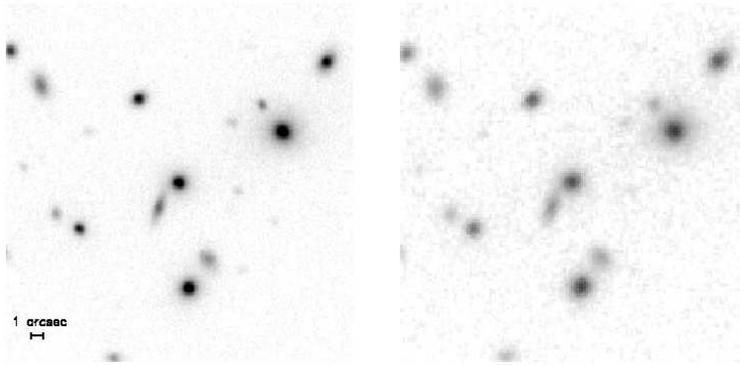}}
\end{center}
\caption[]{ Image of the central region of \cl0048 in the high
resolution (left panel) and in the standard resolution (right panel)
I-band images. The images are normalized to the same intensity scale.
Note the spatial scale in the lower-left panel.
\label{CLIMAG}
}
\end{figure}
New I-band images of the cluster of galaxies \cl0048 have been
collected at the ESO Very Large Telescope (VLT) during August 2001
with the FORS2 instrument, by using the high resolution observing mode
(pixel scale $\mathrm{0.1 ''/pxl}$). The nights were photometric with
excellent seeing conditions ($\mathrm{ 0.3'' < FWHM < 0.4''}$).  The
data consist of three pointings, for each of which we obtained five
dithered exposures of $360~\mathrm{sec}$ each.  Data reduction was
performed as described in LMI03, by using IRAF and Fortran routines
developed by the authors. Images were bias subtracted and corrected
for flat-field by using twilight sky exposures. The frames of each
night were median combined in order to obtain a super-flat frame, that
was used to improve the accuracy of the flat correction (at better
than $\sim 0.5 \%$) and to fully remove the fringing pattern in the
images (see LMI03 for details). The frames were registered by integer
shifts and combined by using the IRAF task IMCOMBINE. The photometric
calibration was performed into the Johnson-Kron-Cousins system by
using comparison standard fields from Landolt~(\cite{LAN92}) observed
during each night. The accuracy on the photometric calibration amounts
to $\mathrm{\sim 0.005~mag}$.  In Fig.~\ref{CLIMAG}, we show, as
example of the final quality of the image, the central region of the
cluster. The image is compared with that obtained in the I band under
more ordinary seeing conditions ($\mathrm{FWHM \sim 0.8''}$) with the
standard resolution observing mode (pixel scale of $\mathrm{0.2
''/pxl}$, see LMI03).

\section{PSF of the FORS2 I-band images.}
\begin{figure}
\begin{center}
\resizebox{9cm}{4.5cm}{\includegraphics{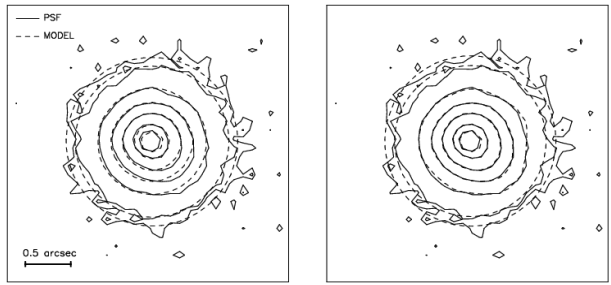}}
\end{center}
\caption[]{ Modelling of the I-band PSF for \cl0048. The contour plots
refer to a star in the cluster field (continuous lines) and to the PSF
models (dashed curves). Left and right panels show the symmetric and
asymmetric models, respectively.
\label{psfmod}
}
\end{figure}

The PSF of the FORS2 images was modelled by using a multi-Gaussian
expansion, taking into account the effect of pixel convolution (see
LBM02).  Stars were selected according to the stellar index (SG) of
SExtractor (Bertin \& Arnout~\cite{BeA96}). We considered only objects
with $I < 20$ and $\mathrm{SG} > 0.95$, in order to obtain a
reasonable accuracy in the fitting and to minimize the contamination
by extended sources. In Fig.~\ref{psfmod}, we show the contour plot
for a star in the cluster field and the corresponding circular model,
obtained as described in LBM02. It is evident that the star is
asymmetric in a direction at $\sim -45$ degrees with respect to the
horizontal axis. This effect can be explained by the fact that five dithered
exposures were available for each pointing and the seeing was not
exactly the same for each of them. Since the images were combined by
integer shifts, the seeing difference can easily account for the PSF
asymmetry. We found that this effect can introduce a significant bias
($10 $-- $15 \%$) in the measurement of the structural parameters and,
therefore, we considered more suitable models of the PSF, by using the
rotation angle of each Gaussian as a free parameter in the fit and by
introducing another free parameter to describe the deviations of the
PSF isophotes from the circular shape. In Fig.~\ref{psfmod}, we see
that the asymmetric model matches very well the star isophotes.  For
each image, we fitted first each star separately, in order to
investigate possible variations of the PSF across the chip, and then,
since no significant dependence on the position was found, we
constructed the PSF model by fitting all the stars simultaneously.

\section{Surface brightness profiles.}

The UV, OPT and NIR surface brightness profiles of the
$\mathrm{N=273}$ galaxies of \a209, \ac118 \, and \cl0048 \, are shown
in Figs. C.1, C.2, and 
C.3\footnote{Figures are available in electronic form at
http://www.edpsciences.org.},
respectively. The 1D profiles are plotted down to a signal to noise
ratio $\mathrm{S/N=1}$.  For each galaxy, we sampled the observed,
fitted, and de-convolved images along ellipses with center
coordinates, axis ratio, and position angle given by the
two-dimensional fit, and derived the mean surface brightness
$\mathrm{\mu(r)}$ as a function of the equivalent radius $\mathrm{r}$
of each ellipse. We considered only the pixels not masked in the 2D
fit and excluded ellipses with a low fraction ($<30\%$) of unmasked
pixels.  For the observed profiles, the error bars on
$\mathrm{\mu(r)}$ were computed by taking into account the surface
brightness fluctuations along the ellipses as well as the local
background noise of each image. The profiles of galaxies belonging to
multiple systems were obtained by first subtracting the 2D models of
the companions from the observed galaxy image.

\end{document}